\documentclass[prb,singlecolumn,preprintnumbers,amsmath,amssymb,letterpaper]{revtex4}

\usepackage{graphicx}
\usepackage{latexsym}
\usepackage{amsmath}
\usepackage{graphics}
\usepackage{amssymb}
\usepackage{layout}
\usepackage{verbatim}
\usepackage{amsfonts,epsfig}
\usepackage{slashed}
\usepackage{feynmp}
\usepackage{float}
\usepackage{color}

\newcommand{\beq}{\begin{equation}}
\newcommand{\eeq}{\end{equation}}
\newcommand{\bea}{\begin{eqnarray}}
\newcommand{\eea}{\end{eqnarray}}
\newcommand{\nn}{\nonumber}

\newcommand{\x}{\xi}

\begin{document}

\unitlength = 1mm

\title{Robust quantum control using smooth pulses and topological winding}

\author{Edwin Barnes$^{1,2}$, Xin Wang$^{1}$, and S.~Das Sarma$^{1,2}$}

\affiliation{$^{1}$Condensed Matter Theory Center, Department of Physics, University
of Maryland, College Park, MD 20742, USA\\
$^{2}$Joint Quantum Institute, University of
Maryland, College Park, MD 20742, USA}

\begin{abstract}
The greatest challenge in achieving the high level of control needed for future technologies based on coherent quantum systems is the decoherence induced by the environment.  Here, we present an analytical approach that yields explicit constraints on the driving field which are necessary and sufficient to ensure that the leading-order noise-induced errors in a qubit's evolution cancel exactly. We derive constraints for two of the most common types of noise that arise in qubits: slow fluctuations of the qubit energy splitting and fluctuations in the driving field itself. By theoretically recasting a phase in the qubit's wavefunction as a topological winding number, we can satisfy the noise-cancelation conditions by adjusting driving field parameters without altering the target state or quantum evolution. We demonstrate our method by constructing robust quantum gates for two types of spin qubit: phosphorous donors in silicon and nitrogen-vacancy centers in diamond.
\end{abstract}

\maketitle

Quantum-based devices are anticipated to serve as the foundation for a new wave of technology capable of performing tasks far beyond the reach of present day electronics. At the heart of this expectation is the demand for microscopic quantum systems that can be reliably manufactured, isolated from their environment, and controlled with very high precision. Residual effects from the environment are of course inevitable, especially for solid state devices, where decoherence stems from a variety of sources, including charge noise,\cite{Maune_Nature12,Shulman_Science12} nuclear spin fluctuations,\cite{Petta_Science05,Shulman_Science12,vanderSar_Nature12} stray magnetic fields,\cite{Petta_Science05} quasiparticle poisoning,\cite{Pop_Nature14} etc. Some level of environmental disturbance is acceptable provided these effects are not so strong as to destroy the coherence of the system before it has completed its task.\cite{Shor_PRA95} It has been known for several decades that a crucial ingredient in achieving this tolerance threshold is the use of carefully designed control protocols capable of dynamically correcting for the effects of noise.\cite{Hahn_PR50,Carr_Purcell,Wimperis_JMR94,Merrill_Wiley14} Such methods are particularly effective in the case of a non-Markovian environment that induces fluctuations in system properties that are slow compared to the control timescales.

The search for robust control fields has been carried out over the last several decades, originating in the field of NMR and branching into newer fields such as quantum computing and nanoscale devices. Dynamical control techniques for preserving the state of an idle two-level system coupled to an environmental bath (e.g. spin echo \cite{Hahn_PR50}) have in fact been known for more than 60 years. In the context of quantum computing, considerable progress has been made in recent years in developing more sophisticated control protocols that extend the lifetime of a quantum state, an important step toward constructing quantum memory resources.\cite{Uhrig_PRL07,Witzel_PRL07,Du_Nature09} However, for the purposes of quantum information processing, it is also necessary to correct errors while a computation is being performed; this is a much more challenging objective, and it is our goal in this work. Several approaches have been pursued previously to create controls that execute a desired quantum evolution while simultaneously combatting noise using either numerical or analytical methods.\cite{Goelman_JMR89,Khodjasteh_PRL10,vanderSar_Nature12,Wang_NatComm12,Kestner_PRL13,Wang_PRA14,Khodjasteh_PRA12,Kabytayev_PRA14,Soare_NP14,Fauseweh_PRA12,Noebauer_arXiv14} Numerical methods cannot easily distinguish local and global extrema in a cost function and can be difficult to use depending on the number and nature of physical constraints present in a system of interest.\cite{Wang_NatComm12,Kestner_PRL13} Analytical methods suffer from the problem that very few analytical solutions to the time-dependent Schr\"odinger equation are known,\cite{Barnes_PRL12} a fact which leads to proposals involving sequences of idealized control pulses, such as delta functions or square waveforms, that are often neither optimal nor easily implemented in real experimental setups. Replacing such waveforms with smoother shapes such as Gaussians\cite{Soare_NP14} can make them easier to generate in real systems, but the fact remains that using a preselected pulse shape repeatedly provides few tunable parameters and leads to unnecessarily long control sequences which may be impractical depending on the physical system in question.

In this work, we present a general solution to the quantum control problem for a two-level system. We develop an analytical approach to constructing robust dynamical control protocols that yields an unlimited number of smoothly varying, experimentally feasible driving fields for a given task. Unlike previous analytical methods, we do not preselect a particular waveform to serve as the building block of composite sequences, but rather develop a formalism that systematically generates optimal waveforms. We consider a two-level Hamiltonian of the following form:
\beq
H=\left(\begin{matrix} \Omega(t) & \beta \cr \beta & -\Omega(t) \end{matrix}\right),\label{defofH}
\eeq
where $\beta$ can be thought of as the qubit energy splitting and $\Omega(t)$ is the driving field. This Hamiltonian describes several types of qubit. For example, in the case of
singlet-triplet spin qubits,\cite{Levy.02,Petta_Science05,Maune_Nature12,Shulman_Science12,Wu_PNAS14} $\Omega(t)$ represents a time-dependent exchange coupling between two electron spins, and $\beta$ is a magnetic field gradient. On the other hand, for
spin qubits driven by monochromatic laser or microwave fields, Eq.~(\ref{defofH}) is the Hamiltonian in the rotating frame of the driving field, with $\Omega(t)$ determined by
its power, and $\beta$ is its detuning relative to the qubit's resonance frequency.\cite{Economou_PRL07,Greilich_NP09,Pla_Nature12,Rong.14} Interactions between the qubit and a non-Markovian
environment can induce slow fluctuations in both $\Omega(t)$ and $\beta$ such that $\beta=\beta_0+\delta\beta$ and $\Omega(t)=\Omega_0(t)+g(t)\delta\epsilon$, where $\delta\beta$ and $\delta\epsilon$ are
unknown stochastic variations that are independent of each other and constant during the application of $\Omega(t)$, and $g(t)$ is a function which generally depends on $\Omega_0(t)$ and on the nature of the noise. Our goal is
to choose a form for $\Omega_0(t)$ such that the evolution operator $U(t)$ generated by Eq.~(\ref{defofH}) achieves a target value $U(t_f)$ at some time $t_f$ and where $U(t_f)$ is independent of $\delta\beta$ and $\delta\epsilon$ to first order in these fluctuations. While finding all possible driving fields $\Omega_0(t)$ that achieve this may seem like an impossible task, we show that it proves to be surprisingly tractable. Below, we obtain a general solution to this problem by deriving a set of constraints which any robust control field must obey and showing how these can be solved systematically.

The starting point of our method is a recently proposed formalism for generating forms of $\Omega(t)$ for which the corresponding Schr\"odinger equation can be solved exactly.\cite{Barnes_PRL12,Barnes_PRA13,Garanin_EPL02} The basic idea is to parameterize both the driving field $\Omega(t)$ and the evolution operator $U(t)$ in terms of a single function denoted by $\chi(t)$. In particular, the driving field can be expressed as
\beq
\Omega(t)=\frac{\ddot\chi}{2\sqrt{\beta^2-\dot\chi^2}}-\sqrt{\beta^2-\dot\chi^2}\cot(2\chi).\label{Omegafromchi}
\eeq
A similar expression for $U(t)$ in terms of $\chi(t)$ along with a brief description of the formalism can be found in the Supplementary Information. The main result of Ref.~[\onlinecite{Barnes_PRA13}] is that any choice of $\chi(t)$ obeying the inequality $|\dot\chi|\le|\beta|$ yields an analytical expression for the evolution $U(t)$ generated by the $\Omega(t)$ determined from Eq.~(\ref{Omegafromchi}), where the inequality enforces the unitarity of $U(t)$.

In the present context of designing robust controls, the utility of the $\chi(t)$ formalism is that it allows us to trace how fluctuations in the Hamiltonian give rise to fluctuations in the evolution operator. In particular, the fluctuations $\delta\beta$ and $\delta\epsilon$ will induce time-dependent fluctuations of $\chi$: $\chi(t)=\chi_0(t)+\tfrac{\delta\chi(t)}{\delta\beta}\delta\beta+\tfrac{\delta\chi(t)}{\delta\epsilon}\delta\epsilon$. In the Supplementary Information, we show that, remarkably, the fluctuations in $\chi(t)$ can be calculated exactly analytically:
\beq
\frac{\delta\chi(t)}{\delta\beta}=\frac{2}{\beta_0}\left\{\frac{1}{8}\sin[4\chi_0(t)]+\hbox{Re}\left[e^{-2i\xi_0(t)}\int_0^tdt'\dot\chi_0(t')\sin^2[2\chi_0(t')]e^{2i\xi_0(t')}\right]\right\},\label{chifluca}
\eeq
\beq
\frac{\delta\chi(t)}{\delta\epsilon}=\frac{1}{2}\hbox{Im}\left[e^{-2i\xi_0(t)}\int_0^tdt'\sin[2\chi_0(t')]g(t')e^{2i\xi_0(t')}\right],\label{chiflucb}
\eeq
where $\xi_0(t)=\int_0^tdt'\sqrt{\beta_0^2-\dot\chi_0(t')^2}\csc[2\chi_0(t')]$ is a phase appearing in $U(t)$. Since the evolution $U(t)$ is a functional of $\chi(t)$, we can use Eqs.~(\ref{chifluca}) and (\ref{chiflucb}) to derive the corresponding variations of $U(t)$ due to noise. We can then construct noise-resistant driving fields by requiring these variations to vanish at the final time $t=t_f$; this imposes constraints on $\chi_0(t)$, the solutions to which can then be input into Eq.~(\ref{Omegafromchi}) to obtain forms of $\Omega_0(t)$ that implement robust control.

A challenge of the strategy we have just outlined is that it does not include a means of fixing the net evolution $U(t_f)$ to a target value. For example, we could solve the constraints on $\chi_0(t)$ by picking an ansatz for this function that includes free parameters that can be adjusted until the constraints are satisfied. As we tune these parameters, though, we must also ensure that $U(t_f)$ does not vary, and this is made difficult by the presence of the phase $\xi_0(t)$ in $U(t)$; the fact that this phase is an integral of a complicated nonlinear expression involving $\chi_0(t)$ makes it challenging to hold $\xi_0(t_f)$ fixed as parameters in $\chi_0(t)$ are varied.

We circumvent this formidable problem by observing that the invariance of $\xi_0(t_f)$ under parameter variations is tantamount to saying that this phase is a topological winding number. In this point of view, $\xi_0(t)$ is proportional to the phase of a complex function which traces a contour in the complex plane that winds one or more times around the origin as time evolves from $t=0$ to $t=t_f$. Changing parameters in $\chi_0(t)$ deforms this contour but preserves the winding number provided the contour does not cross the origin. Thus, the quantization of this topological winding number enables us to fix the target evolution while eliminating leading-order unknown errors in the qubit evolution.

\begin{figure}
\begin{center}
\includegraphics[width=\columnwidth]{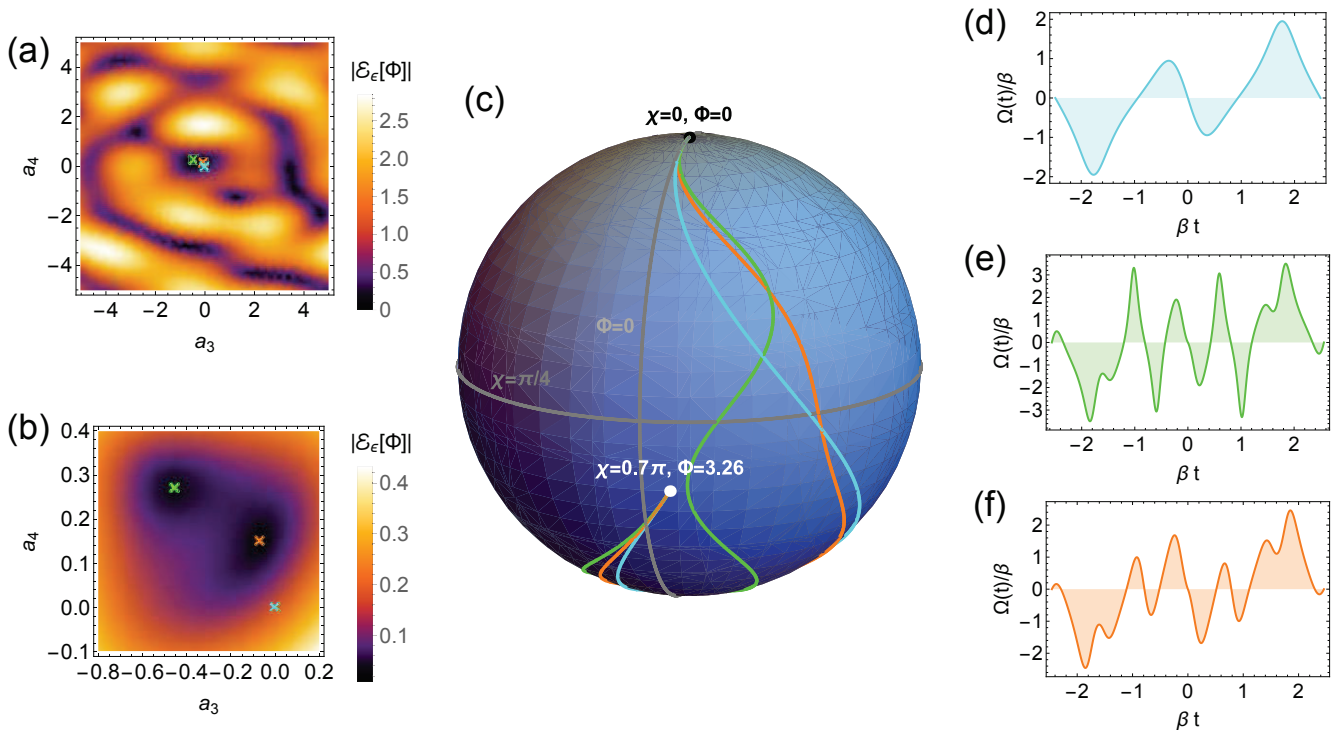}
\caption{\label{fig:fig1} (a) The error potential associated with $\delta\epsilon$-noise with $g(t)=\Omega_0(t)$ for an ansatz of the form $\Phi(\chi)=a_1\chi^2+a_2\chi^3+a_3\sin^3(4\pi\chi/\phi)+a_4\sin^3(8\pi\chi/\phi)$ with $\phi=2.8\pi$, $a_1=0.74$, $a_2=-0.18$, corresponding to a rotation about the axis $\sqrt{3}\hat x+\hat y$ in the $xy$ plane. The parameter regions where the first-order error in the evolution operator vanishes are shown in black. (b) A zoom-in on the central region of (a) revealing two points where the error vanishes (marked with green and orange). (c) Geometrical representation of control fields as curves on the surface of a sphere which extend from the north pole to a final point determined by the target evolution. The green and orange curves correspond to the two points of vanishing error shown in (b), while the cyan curve corresponds to the point $a_3=a_4=0$ at which the error is nonzero. The lengths of the curves give the durations of the respective control fields. (d-f) The control fields for each of the curves shown in (c). Each control field implements the same rotation in approximately the same time. The driving fields shown in (e) and (f) dynamically cancel the $\delta\epsilon$ error, while that shown in (d) does not.}
\end{center}
\end{figure}

In practice, we implement this idea by expressing the phase $\xi_0(t)$ as a functional of $\chi_0(t)$: $\xi_0(t)=\Phi[\chi_0(t)]$. This allows us to control the final value of the phase $\xi_0(t_f)$ directly from the function $\Phi(\chi)$ without fixing $\xi_0(t)$ itself, which would be far too restrictive. By equating the integrand of $\xi_0(t)$ to $\dot\chi_0\Phi'[\chi_0]$, we see that we can reproduce $\chi_0(t)$ from $\Phi(\chi)$ through the formula
\beq
\beta t=\int_0^{\chi_0(t)}d\chi\sqrt{1+\Phi'(\chi)^2\sin^2(2\chi)}.\label{chi0fromPhi}
\eeq
Thus, for a given $\Phi(\chi)$, we can obtain the driving field $\Omega_0(t)$ and complete time-dependence of the evolution operator by first performing the integral in Eq.~(\ref{chi0fromPhi}) and inverting the result to find $\chi_0(t)$. Moreover, as shown in the Supplementary Information, we can convert the noise-cancelation constraints derived earlier for $\chi_0(t)$ into a general set of constraints on $\Phi(\chi)$. For example, in the case of time-antisymmetric driving fields ($\Omega(-t)=-\Omega(t)$ for controls applied from $t=-t_f$ to $t=t_f$), the constraints for canceling $\delta\beta$-noise and $\delta\epsilon$-noise are respectively
\beq
{\cal E}_\beta[\Phi]\equiv\sin(\phi)+8e^{-2i\Phi(\phi/4)}\int_0^{\phi/4} d\chi \sin^2(2\chi)e^{2i\Phi(\chi)}=0,\label{noiseconstraintsa}
\eeq
\beq
{\cal E}_\epsilon[\Phi]\equiv\int_0^{\phi/4} d\chi \sin(2\chi)\sqrt{1+[\Phi'(\chi)]^2\sin^2(2\chi)}\tilde g(\chi)e^{2i\Phi(\chi)}=0,\label{noiseconstraintsb}
\eeq
where $\tilde g(\chi(t))=g(t)$, and $\phi$ is the target rotation angle. We can visualize the solution space of these constraints by choosing an ansatz for $\Phi(\chi)$ that contains adjustable parameters and then plotting $|{\cal E}[\Phi]|$ as a function of these parameters. An example of such an ``error potential" is shown in Fig.~\ref{fig:fig1}(a) for an ansatz containing two free parameters. The points in parameter space where the error potential vanishes yield driving fields that implement robust quantum control. One additional constraint for each type of noise must be satisfied by $\Phi(\chi)$ for more general driving fields (see Supplementary Information). If $\Phi(\chi)$ satisfies both (\ref{noiseconstraintsa}) and (\ref{noiseconstraintsb}), the corresponding evolution will be immune to both types of error. One can also suppress pulse timing errors by imposing constraints on the initial and final values of the higher derivatives of $\Phi(\chi)$, which produces a flattening of the tails of the pulse (see Supplementary Information).

The expression on the right-hand side of Eq.~(\ref{chi0fromPhi}) can be graphically interpreted as the length of a curve lying on the surface of a sphere parameterized by polar angle $\chi/2$ and azimuthal angle $\Phi/2$. This reveals an underlying geometrical picture in which the driving field is represented as a string on the sphere's surface which extends from the north pole ($\chi=0$) down to a point $\chi=\phi/4$, $\Phi=\xi_0(t_f)$ determined by the target evolution $U(t_f)$. Examples are shown in Fig.~\ref{fig:fig1}(c), with the corresponding driving fields displayed in Fig.~\ref{fig:fig1}(d-f). As illustrated in Fig.~\ref{fig:fig1}, the noise-cancelation constraints generally admit multiple solutions, translating to a collection of different strings which all start and end at the same points. Eq.~(\ref{chi0fromPhi}) indicates that the total duration of the control field is given by the length of the string: $t_f=\tfrac{1}{\beta}\int_0^{\phi/4}d\chi\sqrt{1+\Phi'(\chi)^2\sin^2(2\chi)}$. This observation shows that functions $\Phi(\chi)$ which minimize this expression yield the control fields that generate the fastest possible target evolutions. In the context of robust quantum control, this minimization should be performed over the set of solutions to the noise-cancelation constraints.

\begin{figure}
\begin{center}
\includegraphics[width=0.8\columnwidth]{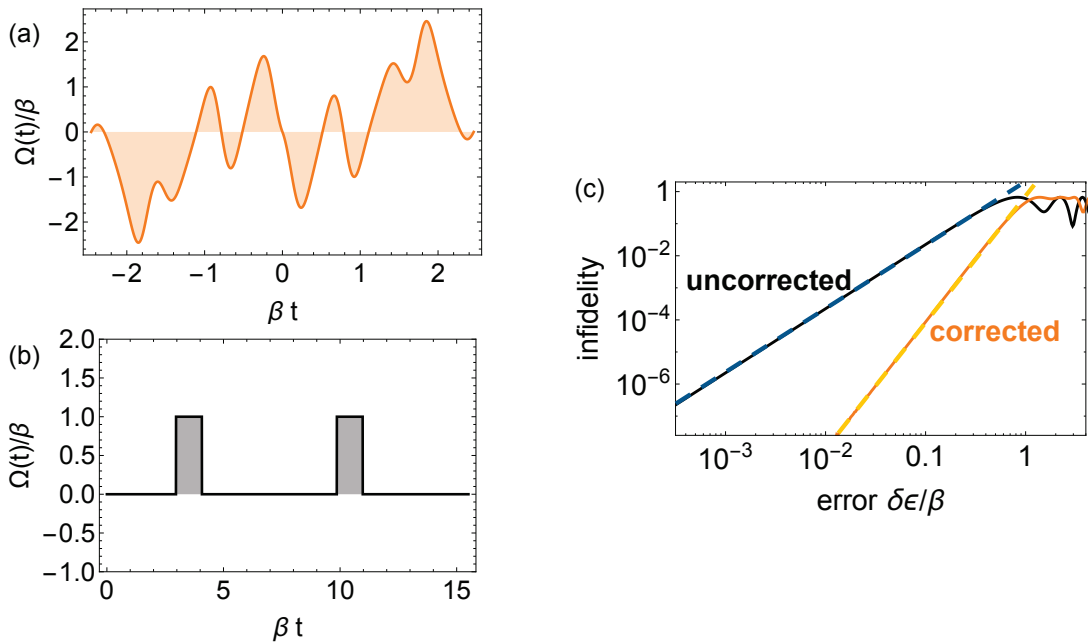}
\caption{\label{fig:dJfig} (a) Pulse from Fig.~\ref{fig:fig1}(f) designed to self-correct for driving field errors ($\delta\epsilon$-errors). This pulse implements a rotation about the axis $\sqrt{3}\hat x+\hat y$ by angle $2.8\pi$. In the context of electron spin qubits in silicon, $\Omega(t)$ is the envelope and $\beta$ is the detuning of an external microwave field. Typical microwave generators produce waveforms obeying $\Omega(t)\lesssim3$MHz,\cite{Pla_Nature12} in which case we should choose $\beta\approx1$MHz, yielding a 5$\mu$s pulse. (b) An ordinary pulse that implements the same rotation as the pulse in (a) but which does not satisfy the error-cancelation constraint in Eq.~(\ref{noiseconstraintsb}). (c) Comparison of the infidelities incurred by the control fields of (a) and (b). The orders of magnitude reduction in the infidelity and the change in the slope of the corrected curve relative to the uncorrected demonstrate first-order error cancelation. Blue and orange dashed lines are quadratic and quartic fits, respectively: 2.2$(\delta\epsilon/\beta)^2$, 0.9$(\delta\epsilon/\beta)^4$.}
\end{center}
\end{figure}

We demonstrate our method by applying it to two types of solid state qubits which are currently at the forefront of quantum technology research. The first type of qubit is comprised of the two spin states of an electron confined to a phosphorous donor in silicon,\cite{Kane_Nature98,Morello_Nature10,Tyryshkin_NatMat11,Pla_Nature12,Wolfowicz_NatNano13} where one of the primary manifestations of noise stems from power fluctuations in the waveform generators used to implement the control fields.\cite{Pla_Nature12, Pla.13} Provided that these power fluctuations are slow compared to the duration of the applied field, we can model this as $\delta\epsilon$-noise where the function $g(t)$ characterizing the noise is proportional to the intended field $\Omega_0(t)$. We use the ansatz for $\Phi(\chi)$ given in Fig.~\ref{fig:fig1} which yields rotations about a particular axis in the $xy$ plane (see Supplementary Information for a universal set of robust quantum gates). To demonstrate the cancelation of noise, we show the infidelity as a function of the noise strength in Fig.~\ref{fig:dJfig}. For comparison, we have also included the infidelity incurred by an ordinary piecewise square pulse that implements the same rotation. The orders of magnitude reduction in the infidelity and the change in slope of the curve clearly demonstrate the cancelation of the leading-order errors in the target evolution.

\begin{figure}
\begin{center}
\includegraphics[width=0.8\columnwidth]{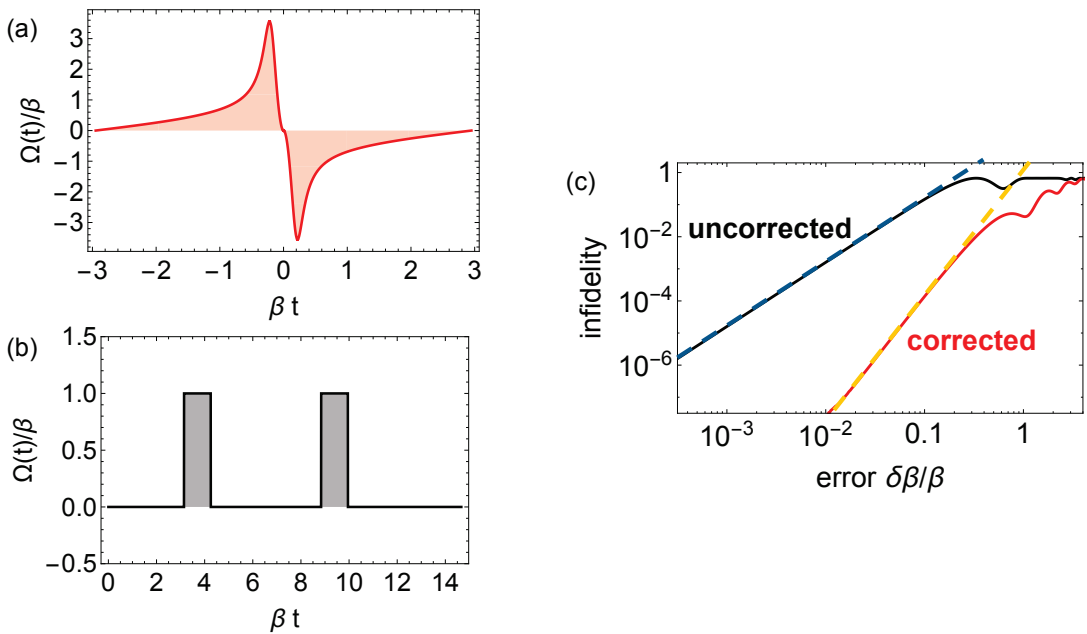}
\caption{\label{fig:dhfig} (a) Driving field derived from choosing $\Phi(\chi)=4\chi-\sin(4\chi)$. This pulse implements a $\pi$ rotation about the axis $\hat x-8\hat y$ while canceling $\delta\beta$-noise, which represents hyperfine noise in the case of NV centers in diamond. In this context, we can interpret $\Omega(t)$ as the amplitude of a microwave pulse with detuning $\beta$. For a pulse of maximum amplitude 20MHz, choosing $\beta=5$MHz leads to a pulse duration of 1.2$\mu$s. (b) An ordinary pulse that implements the same rotation as (a) but without built-in error suppression. (c) A comparison of the infidelities incurred by the driving fields shown in (a) and (b) exhibits the noise cancelation effected by pulse (a). Blue and orange dashed lines are quadratic and quartic fits, respectively: 15$(\delta\beta/\beta)^2$, 1.5$(\delta\beta/\beta)^4$.}
\end{center}
\end{figure}

The second example we consider is a spin qubit in a nitrogen-vacancy center in diamond.\cite{Grinolds_NatPhys11,vanderSar_Nature12,Rong.14} In this case, a leading source of noise is fluctuations in the qubit energy splitting due to hyperfine interactions with neighboring nuclear spins.\cite{vanderSar_Nature12,Rong.14} These fluctuations are typically very slow and naturally modeled in terms of $\delta\beta$-noise. We can again use an ansatz like that given in Fig.~\ref{fig:fig1} and tune parameters to satisfy the noise-cancelation condition. Details along with parameters for a complete set of universal gates are given in the Supplementary Information. Here, we note that ${\cal E}_\beta[\Phi]$ can also be made to vanish exactly when $\phi=n\pi$ for some integer $n$ by choosing $\Phi(\chi)=[4\chi-\sin(4\chi)+F(4\chi-\sin(4\chi))]/n$, where $F(\theta+n\pi/2)=F(\theta)$ is any periodic function with period $n\pi/2$. One of the corresponding driving fields which produces a $\pi$ rotation about an axis in the $xy$ plane is shown in Fig.~\ref{fig:dhfig}(a). A striking reduction in noise relative to the performance of a generic control field (see Fig.~\ref{fig:dhfig}(b)) is revealed in a comparison of the respective infidelities, shown in Fig.~\ref{fig:dhfig}(c).

The results presented here show that analytical methods based on a deep theoretical analysis of qubit dynamics can be a powerful tool in developing experimentally feasible robust quantum controls involving the application of smooth practical external pulses. Future work will include further optimization in terms of minimizing the control durations, including additional driving terms in the Hamiltonian, allowing for non-static noise with a well-defined power spectrum, and extending the approach to multi-level quantum systems. We anticipate that such methods will play an important role in overcoming the decoherence problem in microscopic quantum systems. In particular, we believe that the theoretical techniques presented here will be essential in reducing errors in solid state quantum computing architectures down to the level of the quantum error correction threshold so that scalable quantum information processing may become feasible in the laboratory.

{\bf Acknowledgments}  This work is supported by LPS-NSA-CMTC and IARPA-MQCO.

\appendix

\section{Analytically solvable driving fields}
As shown in Ref.~\onlinecite{Barnes_PRA13}, the dynamical evolution operator for a general, time-dependent two-level quantum system described by the Hamiltonian
\beq
H=b_x(t)\sigma_x+b_y(t)\sigma_y+b_z(t)\sigma_z,\label{ham}
\eeq
can be written in the form
\beq
U=\left(\begin{matrix}u_{11} & -u_{21}^*\cr u_{21} & u_{11}^*\end{matrix}\right), \qquad |u_{11}|^2+|u_{21}|^2=1,\label{defofU}
\eeq
where the explicit $u_{11}$, $u_{21}$ and driving fields are
\bea
u_{11}&=&\cos\chi e^{i\xi_--i\varphi/2},\quad u_{21}=i\eta\sin\chi e^{i\xi_++i\varphi/2},\label{evolfromchi}\\
\xi_\pm&=&\int_0^tdt'\beta\sqrt{1-{\dot\chi^2\over\beta^2}}\csc(2\chi)\pm{1\over2}\arcsin\left({\dot\chi\over\beta}\right)\pm\eta{\pi\over4},\nn\\
b_x&=&\beta\cos\varphi,\quad b_y=\beta\sin\varphi,\nn\\
b_z&{=}&{\ddot\chi{-}\dot\chi\dot\beta/\beta\over2\beta\sqrt{1{-}\dot\chi^2/\beta^2}}{-}\beta\sqrt{1{-}\dot\chi^2/\beta^2}\cot(2\chi){+}\frac{\dot\varphi}{2}.\label{bfromchi}
\eea
$\beta(t)$, $\varphi(t)$, and $\chi(t)$ are three auxiliary functions that allow us to more easily determine how the properties of $H$ influence the behavior of $U$. The main result of Ref.~\onlinecite{Barnes_PRA13} is that any solution to the time-dependent Schr\"odinger equation corresponds to some choice of these three auxiliary functions for which the inequality $|\dot\chi|\le|\beta|$ is satisfied. The initial conditions $u_{11}(0){=}1$, $u_{21}(0){=}0$ imply $\chi(0){=}0$, and $\dot\chi(0){=}{-}\eta\beta(0)$ ensures $b_z(0)$ is finite, where $\eta{=}{\pm}1$. In the most general case of driving along all three axes, all three auxiliary functions have nontrivial time dependence. In this work, we are only interested in the case of single-axis driving, for which we have $\varphi=0$ and $\beta$ is a constant. In this case, the Hamiltonian reduces to the form
\beq
H=\beta\sigma_x+\Omega(t)\sigma_z,\label{hamii}
\eeq
where we have defined $b_z(t)=\Omega(t)$. This Hamiltonian is in a form that is relevant for several types of experimentally relevant two-level quantum systems, for example singlet-triplet spin qubits in semiconductor quantum dots.\cite{Petta_Science05}

In many types of qubits, particularly those used in electron spin resonance or nuclear magnetic resonance experiments, the effective qubit Hamiltonian can be expressed as
\beq
H_{qubit}=\left(\begin{matrix} E/2 & \Omega(t)e^{-i\omega t}\cr \Omega(t)e^{i\omega t} & -E/2\end{matrix}\right),\label{Htransversedriving}
\eeq
where $E$ is the qubit energy splitting and $\Omega(t)e^{i\omega t}$ represents a monochromatic pulse with envelope $\Omega$ and frequency $\omega$. The simplest way to adapt our formalism to this case is to first perform the following transformation to the rotating frame:
\beq
T=e^{-i\tfrac{\omega t}{2}\sigma_z}e^{-i\tfrac{\pi}{4}\sigma_y},
\eeq
yielding
\beq
H_{rot}=\left(\begin{matrix}\Omega(t) & (\omega-E)/2\cr (\omega-E)/2 & -\Omega(t)\end{matrix}\right).
\eeq
This rotating-frame Hamiltonian has the form of Eq.~(\ref{hamii}) with $\beta=(\omega-E)/2$. In the context of NV centers or electron spin qubits in phosphorous donors in silicon, fluctuations in $\beta$ result from fluctuations in $E$, which in turn are caused by Overhauser noise arising from nuclear spins in the environment. In both these contexts, as well as in the case of nuclear spin qubits in phosphorous donors, noise in $\Omega(t)$ would result from e.g., power fluctuations in the pulse generator. Such fluctuations were identified as a possible cause of nuclear spin control infidelities.\cite{Pla.13}

Although $H_{rot}$ has the same form as Eq.~(\ref{hamii}), it is important to note that we want the evolution operator in the lab frame (i.e., the evolution operator associated with $H_{qubit}$ above) to be the identity matrix at $t=0$. The evolution operators corresponding to $H_{qubit}$ and $H_{rot}$ are related by
\beq
U_{lab}=TU_{rot},
\eeq
so that our initial condition for $U_{rot}$ should then be
\beq
U_{rot}(0)=T^\dagger(0)=e^{i\tfrac{\pi}{4}\sigma_y}=\frac{1}{\sqrt{2}}\left(\begin{matrix} 1 & 1\cr -1 & 1\end{matrix}\right).
\eeq
One way to incorporate this new initial condition would be to simply multiply $U_{rot}$ on the right by $T^\dagger(0)$. This will re-arrange the components of the evolution operator shown in Eq.~(\ref{evolfromchi}) but does not modify the error cancelation procedure we will develop using the Hamiltonian given in Eq.~(\ref{hamii}).

\section{Fluctuations in $\beta$}\label{sec:hnoise}

In this section, we want to consider the situation where the qubit splitting (or external field detuning) $\beta$ exhibits stochastic fluctuations that are constant throughout the duration of the pulse, i.e., $\beta=\beta_0+\delta\beta$, where $\beta_0$ is a known constant and $\delta\beta$ is an unknown constant. We would like to engineer pulses that execute a target evolution while minimizing the errors caused by the unknown variation $\delta \beta$. In order to set up this problem, it turns out to be beneficial to use a new parametrization of the driving field and the corresponding evolution operator. We first derive this new parametrization and then return to the question of how to incorporate stochastic fluctuations into our formalism for generating analytical solutions of two-level quantum dynamics.

The first step in deriving the new parametrization is to transform to a rotating frame in the $x$-basis:
\beq
D_\pm={1\over\sqrt{2}}e^{\pm i\beta t}(u_{11}\pm u_{21}).\label{Dumap}
\eeq
The functions $D_\pm$ then solve the following set of equations which follow from the Schr\"odinger equation for the evolution operator $U$:
\beq
\dot D_\pm=-i\Omega e^{\pm 2i\beta t}D_\mp.\label{Dschrod}
\eeq
These two equations can be solved easily for $\Omega(t)$:
\beq
\Omega^2=-{\dot D_+ \dot D_-\over D_+ D_-},
\eeq
which implies that we may write
\beq
{\dot D_+\over D_+}=-i{\Omega/u},\quad {\dot D_-\over D_-}=-i{\Omega u},
\eeq
where $u=u(t,\beta)$ is an unknown complex function. These two equations are easily solved:
\bea
D_+&=&{1\over\sqrt{2}}e^{-i\int_0^t dt'\Omega(t')/u(t',\beta)},\nn\\
D_-&=&{1\over\sqrt{2}}e^{-i\int_0^tdt'\Omega(t')u(t',\beta)},\label{DpDmfromu}
\eea
where we have imposed the initial conditions $D_\pm(0)=1/\sqrt{2}$.

Equations (\ref{Dschrod}) also imply
\beq
\dot D_+ D_+ e^{-2i\beta t}=\dot D_- D_- e^{2i\beta t}.
\eeq
Plugging Eqs.~(\ref{DpDmfromu}) into this equation yields
\beq
e^{2i\int_0^tdt' \Omega(t')[u(t',\beta)-1/u(t',\beta)]}=u^2(t,\beta)e^{4i\beta t}.\label{uJeqn}
\eeq
Differentiating both sides of this equation with respect to time and performing some algebraic manipulations results in
\beq
\Omega=-i{\dot u+2i\beta u\over u^2-1}.
\eeq
Writing
\beq
u(t,\beta)=\tanh[w(t,\beta)],
\eeq
we have
\beq
\Omega(t)=i\dot w-\beta\sinh(2w).\label{Jfromw}
\eeq
It is straightforward to express $D_\pm$ in terms of $w$:
\bea
D_+&=&{1\over\sqrt{2}}\sinh(w)e^{2i\beta\int_0^tdt'\cosh^2(w(t'))},\nn\\
D_-&=&{1\over\sqrt{2}}\cosh(w)e^{2i\beta\int_0^tdt'\sinh^2(w(t'))}.\label{evolfromw}
\eea
Any choice of $w(t)$ such that the $\Omega(t)$ computed from Eq. (\ref{Jfromw}) is real immediately yields an analytical solution to the Schr\"odinger equation. It is straightforward to find the necessary restriction on $w$. Writing $w=w_r+i w_i$ in Eq. (\ref{Jfromw}), the condition $\hbox{Im}[\Omega]=0$ implies
\beq
\dot w_r=\beta\sin(2w_i) \cosh(2w_r).
\eeq
This equation is easily integrated to give
\beq
w_r(t)=\frac{1}{2}\log\tan\left[\beta\int_0^t\sin(2w_i(t'))+c\right].\label{wrfromwi}
\eeq
In this expression, $c$ is an integration constant determined by the boundary condition $w_r(0)=\tfrac{1}{2}\log\tan c$, which we will leave arbitrary, at least for now. We must impose
\beq
0\le 2\beta\int_0^t dt'\sin(2w_i(t'))+2c\le \pi
\eeq
to ensure that $w_r$ is real. Plugging Eq. (\ref{wrfromwi}) into Eq. (\ref{Jfromw}), we find an expression for $\Omega$ in terms of $w_i$:
\beq
\Omega(t)=-\dot w_i+\beta\cos(2w_i)\cot\left[2\beta\int_0^tdt'\sin(2w_i(t'))+2c\right].
\eeq
This result can in fact be obtained from the method given in Ref.~\onlinecite{Barnes_PRL12} by choosing
\beq
q(t)=\cos\left[2\beta\int_0^tdt'\sin(2w_i(t'))+2c\right].
\eeq
So far, we have derived an alternate algorithm for generating analytical solutions to the Schr\"odinger equation: An analytical solution can be produced by choosing any $w_i(t)$ such that $0\le 2\beta\int_0^t dt'\sin(2w_i(t'))+2c\le \pi$.

We now consider the case where $\beta=\beta_0+\delta\beta$, where $\beta_0$ is a known constant and $\delta\beta$ is a small stochastic variation. Since our formula for $\Omega(t)$, Eq. (\ref{Jfromw}), contains an explicit dependence on $\beta$, we must take care to choose $w(t)$ in such a way that the resulting $\Omega(t)$ obeys $\Omega(t,\beta)=\Omega(t,\beta_0)+{\cal O}(\delta \beta^2)$. This condition is necessary since $\Omega(t)$ cannot itself depend on the stochastic variable $\delta\beta$. As long as we are only concerned with first-order variations in $\delta\beta$, it is sufficient to require the first-order variation of $\Omega(t)$ to vanish. Note that the present analysis could be extended to derive constraints which ensure that higher-order errors in the evolution operator vanish by requiring the corresponding higher-order variations of $\Omega(t)$ to vanish.

We can arrange for the first-order variation of $\Omega(t)$ to vanish by writing $w=w_0+\delta \beta w_1$ and varying Eq. (\ref{Jfromw}) to obtain
\beq
\Omega(t)=i\dot w_0-\beta_0\sinh(2w_0)
+\left[i\dot w_1-\sinh(2w_0)-2\beta_0w_1\cosh(2w_0)\right]\delta\beta+{\cal O}(\delta\beta^2).
\eeq
The first-order variation will vanish if $w_1$ is given by
\beq
w_1(t)=e^{-2i\beta_0\int_0^tdt'\cosh(2w_0(t'))}\bigg[k
-i\int_0^tdt'e^{2i\beta_0\int_0^{t'}dt''\cosh(2w_0(t''))}\sinh(2w_0(t'))\bigg],\label{w1fromw0}
\eeq
where $k$ is an integration constant. What this result means is that we may still use our formalism for constructing analytical solutions even in the presence of the stochastic fluctuation $\delta\beta$. However, unless we arrange for higher-order variations in $\Omega(t)$ to vanish as well, we can only obtain the evolution up to first order in $\delta \beta$. In most physical situations, $\delta \beta\ll \beta_0$, and determining the evolution up to first order in $\delta \beta$ is sufficient. The procedure works by first choosing $w_{0,i}$ to fix the desired driving field, $\Omega(t)$. Choosing $w_{0,i}$ also fixes $w_{0,r}$ and $w_0$, and the latter then determines $w_1$. The corresponding evolution operator to first order in $\delta\beta$ is obtained by using $w=w_0+\delta\beta w_1$ in Eq.~(\ref{evolfromw}).

We have just seen that the $w$-parametrization is useful for removing the $\delta\beta$ dependence from $\Omega(t)$. However, now that we have laid out the necessary steps, it turns out that further simplifications occur if we return to the $\chi$-parametrization. In particular, we may simplify the expression for $w_1$. To see this, begin with the general relation between the two parameterizations:
\beq
\chi(t)=\beta\int_0^tdt'\sin(2w_i(t'))+c.\label{chifromw}
\eeq
Since we wish to impose $\chi(0)=0$, we will set $c=0$ from now on. Using this expression for $\chi(t)$, it is straightforward to show
\beq
w_0=\tfrac{1}{2}\log\tan\chi_0+i\tfrac{1}{2}\sin^{-1}\left(\frac{\dot\chi_0}{\beta_0}\right),
\eeq
so that
\bea
\cosh(2w_0)&=&\csc(2\chi_0)\sqrt{1-\frac{\dot\chi_0^2}{\beta_0^2}}-i\cot(2\chi_0)\frac{\dot\chi_0}{\beta_0},\nn\\
\sinh(2w_0)&=&-\cot(2\chi_0)\sqrt{1-\frac{\dot\chi_0^2}{\beta_0^2}}+i\csc(2\chi_0)\frac{\dot\chi_0}{\beta_0},
\eea
where
\beq
\chi_0(t)=\beta_0\int_0^tdt'\sin(2w_{i,0}(t')).
\eeq
Comparing this result for $\cosh(2w_0)$ with the definition of the function $\x(t)$ defined earlier, we see that
\beq
2\beta_0\int_0^tdt'\cosh(2w_0(t'))=2\xi_0(t)-i\log\sin[2\chi_0(t)],\label{intcoshw0}
\eeq
where here $\xi_0(t)$ is defined in terms of $\chi_0(t)$ (and not the full $\chi(t)$):
\beq
\xi_0(t)=\int_0^tdt'\sqrt{\beta^2-\dot\chi_0(t')^2}\csc[2\chi_0(t')].
\eeq
Substituting this result into Eq.~(\ref{w1fromw0}) and setting the integration constant $k$ therein to zero then yields
\bea
w_1(t)&=&-ie^{-2i\xi_0(t)}\csc[2\chi_0(t)]\int_0^tdt'e^{2i\xi_0(t')}\sin[2\chi_0(t')]\left\{-\cot[2\chi_0(t')]\sqrt{1-\frac{\dot\chi_0^2(t')}{\beta_0^2}} +i\csc[2\chi_0(t')]\frac{\dot\chi_0(t')}{\beta_0}\right\}\nn\\
&=&-\frac{i}{\beta_0}e^{-2i\xi_0(t)}\csc[2\chi_0(t)]\int_0^tdt'e^{2i\xi_0(t')}\left\{-\tfrac{1}{2}\dot\xi_0(t')\sin[4\chi_0(t')]+i\dot\chi_0(t')\right\}\nn\\
&=&-\frac{i}{\beta_0}e^{-2i\xi_0(t)}\csc[2\chi_0(t)]\int_0^tdt'\left\{\frac{d}{dt'}\left[\frac{i}{4}e^{2i\xi_0(t')}\sin[4\chi_0(t')]\right] -\frac{i}{4}e^{2i\xi_0(t')}\frac{d}{dt'}\sin[4\chi_0(t')]+i\dot\chi_0(t')e^{2i\xi_0(t')}\right\}\nn\\
&=&-\frac{i}{\beta_0}e^{-2i\xi_0(t)}\csc[2\chi_0(t)]\left\{\frac{i}{4}e^{2i\xi_0(t)}\sin[4\chi_0(t)]+2i\int_0^tdt'\dot\chi_0(t')e^{2i\xi_0(t')}\sin^2[2\chi_0(t')]\right\}\nn\\
&=&\frac{1}{2\beta_0}\cos[2\chi_0(t)]+\frac{2}{\beta_0}e^{-2i\xi_0(t)}\csc[2\chi_0(t)]\int_0^tdt'\dot\chi_0(t')e^{2i\xi_0(t')}\sin^2[2\chi_0(t')].
\eea
The first-order variation, $\delta w=\delta\beta w_1$, leads to the following variation in $\chi(t)$:
\bea
\delta_\beta\chi(t)&=&\delta\beta\frac{\chi_0(t)}{\beta_0}+2\delta\beta\int_0^tdt'\sqrt{\beta_0^2-\dot\chi_0^2(t')}\hbox{Im}[w_1(t')]\nn\\
&=&\delta\beta\frac{\chi_0(t)}{\beta_0}+2\delta\beta\int_0^tdt'\dot\xi_0(t')\sin[2\chi_0(t')]\hbox{Im}[w_1(t')]\nn\\
&=&\delta\beta\frac{\chi_0(t)}{\beta_0}+\delta\beta\frac{4}{\beta_0}\int_0^tdt'\dot\xi_0(t')\bigg\{\cos[2\xi_0(t')]\int_0^{t'}dt''\dot\chi_0(t'') \sin^2[2\chi_0(t'')]\sin[2\xi_0(t'')]\nn\\&&-\sin[2\xi_0(t')]\int_0^{t'}dt''\dot\chi_0(t'')\sin^2[2\chi_0(t'')]\cos[2\xi_0(t'')]\bigg\}\nn\\
&=&\delta\beta\frac{\chi_0(t)}{\beta_0}+\delta\beta\frac{2}{\beta_0}\bigg\{\sin[2\xi_0(t)]\int_0^tdt'\dot\chi_0(t')\sin^2[2\chi_0(t')]\sin[2\xi_0(t')]\nn\\ &&+\cos[2\xi_0(t)]\int_0^tdt'\dot\chi(t')\sin^2[2\chi_0(t')]\cos[2\xi_0(t')]\bigg\}\nn\\
&&-\delta\beta\frac{2}{\beta_0}\int_0^tdt'\left(\sin^2[2\xi_0(t')]\sin^2[2\chi_0(t')]+\cos^2[2\xi_0(t')]\sin^2[2\chi_0(t')]\right)\dot\chi_0(t')\nn\\
&=&\delta\beta\frac{2}{\beta_0}\bigg\{\frac{1}{8}\sin[4\chi_0(t)]+\sin[2\xi_0(t)]\int_0^tdt'\dot\chi_0(t')\sin^2[2\chi_0(t')]\sin[2\xi_0(t')]\nn\\
&&+\cos[2\xi_0(t)]\int_0^tdt'\dot\chi_0(t')\sin^2[2\chi_0(t')]\cos[2\xi_0(t')]\bigg\}\nn\\
&=&\delta\beta\frac{2}{\beta_0}\left\{\frac{1}{8}\sin[4\chi_0(t)]+\hbox{Re}\left[e^{-2i\xi_0(t)}\int_0^tdt'\dot\chi_0(t')\sin^2[2\chi_0(t')]e^{2i\xi_0(t')}\right]\right\}.\label{deltachih}
\eea
This completes the derivation of Eq.~(3a) of the main text. Below, we will show how to use this final expression for $\delta_\beta\chi$ to construct dynamically corrected qubit operations with smooth analytical pulses.

\section{Fluctuations in $\Omega(t)$}\label{sec:Jnoise}

To see how we can include first-order fluctuations of the form
\beq
\Omega(t)=\Omega_0(t)+\delta\epsilon\frac{d \Omega(t)}{d\epsilon}\equiv \Omega_0(t)+\delta\epsilon g(t),
\eeq
we again start from the $w$-parametrization from Eq.~(\ref{Jfromw}):
\beq
\Omega(t)=i\dot w(t)-\beta\sinh[2w(t)].
\eeq
Writing $w(t)=w_0(t)+\delta\epsilon w_1(t)$, expanding the right hand side to first order in $\delta\epsilon$ and equating all the first-order terms, we find
\beq
g(t)=i\dot w_1-2\beta w_1\cosh(2w_0).
\eeq
This equation is easily solved, with the result
\beq
w_1(t)=e^{-2i\beta\int_0^tdt'\cosh [2w_0(t')]}\left[k-i\int_0^tdt'e^{2i\beta\int_0^{t'}dt''\cosh[2w_0(t'')]}g(t')\right],
\eeq
where $k$ is an integration constant. Using Eq.~(\ref{intcoshw0}), we can rewrite this as
\beq
w_1(t)=e^{-2i\xi_0(t)}\csc[2\chi_0(t)]\left[k-i\int_0^tdt'e^{2i\xi_0(t')}\sin[2\chi_0(t')]g(t')\right].
\eeq
In order to translate this result into a variation $\delta_\epsilon\chi(t)$, we use Eq.~(\ref{chifromw}) which yields
\bea
\delta_\epsilon\chi(t)&=&\delta\epsilon\beta\int_0^tdt'\hbox{Im}[w_1(t')]\cos[2\hbox{Im}[w_0(t')]] =\delta\epsilon\int_0^tdt'\hbox{Im}[w_1(t')]\sqrt{\beta^2-\dot\chi_0(t')^2}\nn\\&=&\delta\epsilon\int_0^tdt'\hbox{Im}[w_1(t')]\dot\xi_0(t')\sin[2\chi_0(t')].
\eea
Using that
\bea
\hbox{Im}[w_1(t)]&=&-\cos[2\xi_0(t)]\csc[2\chi_0(t)]\int_0^tdt'\cos[2\xi_0(t')]\sin[2\chi_0(t')]g(t') \nn\\&&-\sin[2\xi_0(t)]\csc[2\chi_0(t)]\left[k+\int_0^tdt'\sin[2\xi_0(t')]\sin[2\chi_0(t')]g(t')\right],
\eea
and integrating by parts, we have
\bea
\delta_\epsilon\chi(t)&=&\delta\epsilon\frac{1}{2}\bigg\{-\sin[2\xi_0(t)]\int_0^tdt'\cos[2\xi_0(t')]\sin[2\chi_0(t')]g(t') \nn\\&&+\cos[2\xi_0(t)]\left[k+\int_0^tdt'\sin[2\xi_0(t')]\sin[2\chi_0(t')]g(t')\right]\bigg\}\nn\\
&=&\delta\epsilon\frac{1}{2}\left\{k\cos[2\xi_0(t)]+\hbox{Im}\left[e^{-2i\xi_0(t)}\int_0^tdt'\sin[2\chi_0(t')]g(t')e^{2i\xi_0(t')}\right]\right\}.\label{deltachie}
\eea
This completes the derivation of Eq.~(3b) of the main text.

\section{Controlled rotations via topology}\label{sec:topology}

The $\chi$ formalism reviewed above allows us to construct arbitrary rotations by choosing $\chi(t)$ appropriately. However, fixing the target evolution precisely remains challenging because of the fact that the phase $\x(t)=(\x_++\x_-)/2$ appearing in the evolution operator (\ref{evolfromchi}) is given as an integral of a complicated nonlinear expression involving $\chi(t)$:
\beq
\x(t)=\int_0^tdt'\sqrt{\beta^2-\dot\chi^2(t')}\csc[2\chi(t')].
\eeq
Even though we get to choose the form of $\chi(t)$ with only the relatively weak constraint $|\dot\chi|\le|\beta|$ to worry about, it is difficult to predict what values of $\x_f=\x(t_f)$ at the final time $t_f$ can be achieved with a given choice, and ultimately this generally requires numerical evaluation of the integral above. Moreover, if we wish to include tunable parameters in $\chi(t)$ in order to cancel errors and/or improve rotational fidelities, this task is hindered by the complicated nature of the above expression for $\x(t)$ since it will be hard to keep $\x_f$ fixed as such parameters are varied. In this section, we will show how these problems can be circumvented by incorporating the concept of topological winding into our formalism.

The key observation is to notice that the invariance of $\x_f$ under parameter variations in $\chi(t)$ is equivalent to the statement that $\x_f$ is a quantized topological winding number. This quantization can be imposed by choosing $\chi(t)$ in such a way that the integrand in $\x$ is proportional to the derivative of the argument of a complex function $W(t)$:
\beq
\sqrt{\beta^2-\dot\chi^2}\csc(2\chi)=\lambda\partial_t\hbox{arg}[W(t)],\label{chifromW}
\eeq
where $\lambda$ is a real constant that we are free to choose. If $W(t)$ winds around the origin of the complex $W$ plane an integral number ($n$) of times as time evolves from $t=0$ to $t=t_f$, then
\beq
\int_0^{t_f} dt \partial_t\hbox{arg}[W(t)]=2\pi n.
\eeq
The value of $\x_f$ will then be quantized (in units of $2\pi\lambda$) according to
\beq
\x_f=\int_0^{t_f}dt\sqrt{\beta^2-\dot\chi^2}\csc(2\chi)=2\pi n\lambda.
\eeq
This formula shows that we can fix $\x_f$ to be any value we like by choosing $\lambda$ appropriately and by picking a $W(t)$ which exhibits nontrivial winding. Including tunable parameters in $\chi(t)$ is tantamount to including them in $W(t)$; adjusting these parameters will deform the contour traced by $W(t)$ but will leave the winding number $n$ the same so long as the contour does not cross the origin in the process. Once $W(t)$ is chosen, we then solve Eq.~(\ref{chifromW}) to obtain $\chi(t)$, from which the corresponding driving field, $\Omega(t)$, follows from Eq.~(\ref{bfromchi}).

The procedure outlined in the previous paragraph allows us to hold fixed $\x_f$ (and hence the rotation axis and angle) while we adjust parameters in $\chi(t)$. However, it introduces a new difficulty in that the step of solving Eq.~(\ref{chifromW}) can be challenging due to its strongly nonlinear nature. We can simplify this task by expressing $W(t)$ as a function of $\chi(t)$:
\beq
W(t)=Y(\chi(t)).
\eeq
This allows us to rewrite Eq.~(\ref{chifromW}) as
\beq
\dot\chi^2\left\{1+\lambda^2(\hbox{Im}[Y'(\chi)/Y(\chi)])^2\sin^2(2\chi)\right\}=\beta^2.\label{chifromY}
\eeq
Since the term in curly brackets is now purely a function of $\chi$, we may solve this equation by simple integration:
\beq
\beta t=\int_0^{\chi} d\tilde\chi\sqrt{1+\lambda^2(\hbox{Im}[Y'(\tilde\chi)/Y(\tilde\chi)])^2\sin^2(2\tilde\chi)}.\label{tfromchi}
\eeq
Instead of solving a nonlinear differential equation, the task has now been reduced to performing an ordinary integral and inverting the result. To ensure that $\x_f$ is quantized, we now need to choose $Y(\chi)$ to be such that it winds an integral number of times around the origin of the complex $Y$ plane as $\chi$ evolves from $0$ to its final value, $\chi_f$. We can further streamline the determination of $b_z(t)$ by expressing this quantity in terms of $\chi(t)$ using Eq.~(\ref{chifromY}). Explicitly, we find
\beq
\Omega(t)=\frac{\beta F'(\chi)}{2\lambda\hbox{Im}[Y'/Y]\sin(2\chi)}-\beta\lambda F(\chi)\hbox{Im}[Y'/Y]\cos(2\chi),
\eeq
where
\beq
F(\chi)\equiv\left\{1+\lambda^2(\hbox{Im}[Y'(\chi)/Y(\chi)])^2\sin^2(2\chi)\right\}^{-1/2}.
\eeq
We can also express $\x(t)$ as a function of $\chi$:
\bea
\x(t)&=&\lambda\int_0^tdt'\partial_{t'}\hbox{arg}[W(t')]=\lambda\int_0^tdt'\hbox{Im}[\dot W/W]=\lambda\int_0^tdt'\dot\chi(t')\hbox{Im}[Y'(\chi(t'))/Y(\chi(t'))] \nn\\&=&\lambda\int_0^{\chi(t)} d\tilde\chi\hbox{Im}[Y'(\tilde\chi)/Y(\tilde\chi)]=\lambda\hbox{Im}\log[Y(\chi(t))/Y(0)].
\eea

In practice, it turns out to be easier to obtain simpler-looking functions $b_z(t)$ if we work directly with the function
\beq
\Phi(\chi)\equiv \lambda\hbox{Im}\log[Y(\chi)/Y(0)],\label{defofPhi}
\eeq
rather than with $Y(\chi)$. The driving field is fully determined by $\Phi(\chi)$:
\beq
\Omega(\chi)=-\beta\frac{\Phi''(\chi)\sin(2\chi)+4\Phi'(\chi)\cos(2\chi)+2[\Phi'(\chi)]^3\sin^2(2\chi)\cos(2\chi)}{2\left\{1+[\Phi'(\chi)]^2\sin^2(2\chi)\right\}^{3/2}}.\label{bzfromPhi}
\eeq
To obtain the driving field as a function of time, we need to solve the following equation for $\chi$ (see Eq.~(\ref{tfromchi})):
\beq
\beta t=\int_0^{\chi} d\tilde\chi\sqrt{1+[\Phi'(\tilde\chi)]^2\sin^2(2\tilde\chi)}.\label{chifromPhi}
\eeq
The integrand on the right-hand-side can be identified as the line element obtained from the following metric for a 2-sphere:
\beq
ds^2=d\chi^2+\sin^2(2\chi)d\Phi^2,
\eeq
where we view $\chi$ as the azimuthal angle and $\Phi$ as the polar angle. This in turn implies that the duration of the pulse,
\beq
2t_f=\frac{2}{\beta}\int_0^{\chi_f} d\tilde\chi\sqrt{1+[\Phi'(\tilde\chi)]^2\sin^2(2\tilde\chi)},
\eeq
is just the length of the curve defined by $\Phi(\chi)$ on the surface of the 2-sphere (times $2/\beta$).

\section{Derivation of general error-cancelation constraints}

We are now ready to combine all of the ingredients developed in the previous sections to systematically construct rotations that dynamically correct for errors induced by fluctuations in either $\beta$ or $\Omega(t)$. In order to obtain nontrivial results, it is important that we use the unperturbed quantities $\chi_0(t)$, $\beta_0$, and $\xi_0(t)$ in the topological winding formalism developed above.

\subsection{$\beta$ noise}

From the general form of the evolution operator, Eq.~(\ref{evolfromchi}), we see that in the case of a general single-axis driving field, the first-order error due to $\delta\beta$-noise will vanish only if the following three conditions on the first-order variations hold:
\beq
\delta_\beta\chi(t_f)=0,\qquad \delta_\beta\dot\chi(t_f)=\frac{\dot\chi(t_f)}{\beta_0}\delta\beta,\qquad \delta_\beta\xi(t_f)=0.\label{dherror}
\eeq
Combining the topological winding formalism developed in the previous section with our expression for $\delta_\beta\chi(t)$ in the case of fluctuations in $\beta$, Eq.~(\ref{deltachih}), we have
\bea
\delta_\beta\chi(t)&=&\delta\beta\frac{2}{\beta_0}\bigg\{\frac{1}{8}\sin[4\chi_0(t)]+\hbox{Re}\left[e^{-2i\Phi[\chi_0(t)]}\int_0^{\chi_0(t)}d\tilde\chi\sin^2(2\tilde\chi)e^{2i\Phi(\tilde\chi)}\right]\bigg\}.
\eea
We will make this expression slightly more compact by defining the function
\beq
G_\beta(\chi)\equiv \int_0^\chi d\tilde\chi \sin^2(2\tilde\chi)e^{2i\Phi(\tilde\chi)},
\eeq
yielding
\bea
\delta_\beta\chi(t)&=&\delta\beta\frac{2}{\beta_0}\bigg\{\frac{1}{8}\sin[4\chi_0(t)]+\hbox{Re}\left(e^{-2i\Phi[\chi_0(t)]}G_\beta[\chi_0(t)]\right)\bigg\}.\label{deltahchifromPhi}
\eea
Introducing the shorthand notation $\delta\chi_f=\delta\chi(t_f)$, etc., the end-point fluctuations are then given by
\bea
\delta_\beta\chi_f&=&\delta\beta\frac{2}{\beta_0}\left\{\frac{1}{8}\sin(4\chi_f)+\hbox{Re}[e^{-2i\Phi(\chi_f)}G_\beta(\chi_f)]\right\},\nn\\ \delta_\beta\dot\chi_f&=&\delta\beta\frac{\dot\chi_f}{\beta_0}\left\{1+4\Phi'(\chi_f)\hbox{Im}[e^{-2i\Phi(\chi_f)}G_\beta(\chi_f)]\right\},\label{deltahchi}
\eea
and imposing the first two error-cancelation conditions from Eq.~(\ref{dherror}) amounts to requiring
\bea
\sin(4\chi_f)+8\hbox{Re}[e^{-2i\Phi(\chi_f)}G_\beta(\chi_f)]&=&0,\nn\\
\hbox{Im}[e^{-2i\Phi(\chi_f)}G_\beta(\chi_f)]&=&0,\label{canceldherrori}
\eea
which is equivalent to the single complex condition
\beq
\sin(4\chi_f)+8e^{-2i\Phi(\chi_f)}G_\beta(\chi_f)=0.\label{canceldherrorii}
\eeq

Notice that in these conditions, we have omitted the possibility that $\Phi'(\chi_f)=0$, which would have automatically guaranteed that $\delta\dot\chi_f=(\dot\chi_f/\beta_0)\delta\beta$, meaning that we could ignore the second condition in Eq.~(\ref{canceldherrori}). The reason for this is that when $\Phi'(\chi_f)=0$ (which implies $\dot\chi_f=\beta_0$), we do not completely cancel the first-order error in the final evolution operator because of a square root appearing in its first-order variation stemming from the arcsine part of $\xi_\pm$:
\beq
\delta_\beta \left(e^{\pm \tfrac{i}{2}\arcsin(\dot\chi_f/\beta)}\right)=\pm i\left(\delta_\beta\dot\chi_f-\frac{\dot\chi_f}{\beta_0}\delta\beta\right)\frac{e^{\pm \tfrac{i}{2}\arcsin(\dot\chi_f/\beta_0)}}{2\sqrt{\beta_0^2-\dot\chi_f^2}}.
\eeq
If $\dot\chi_f=\beta_0$, then the square root in the denominator vanishes, implying that this part of the first-order variation in the evolution will not vanish in general. The only way to guarantee that the first-order error in the evolution operator is completely removed is to {\it not} impose $\dot\chi_f=\beta_0$.

It remains to analyze the third condition in Eq.~(\ref{dherror}). To calculate $\delta_\beta\xi_f$, we begin with the general expression for $\xi(t)$:
\beq
\xi(t)=\int_0^tdt'\sqrt{\beta^2-\dot\chi^2}\csc(2\chi).
\eeq
The square root inside the integrand can be expanded as follows:
\beq
\sqrt{(\beta_0+\delta\beta)^2-(\dot\chi_0+\delta\dot\chi)^2}\approx\sqrt{\beta_0^2-\dot\chi_0^2} -\frac{\dot\chi_0\delta\dot\chi}{\sqrt{\beta_0^2-\dot\chi_0^2}}+\frac{\beta_0\delta\beta}{\sqrt{\beta_0^2-\dot\chi_0^2}},
\eeq
while the remaining factor becomes
\beq
\csc[2(\chi_0+\delta\chi)]\approx\csc(2\chi_0)-2\delta\chi\cot(2\chi_0)\csc(2\chi_0).
\eeq
We therefore have
\beq
\delta\xi(t)=-2\int_0^tdt'\sqrt{\beta_0^2-\dot\chi_0^2}\cot(2\chi_0)\csc(2\chi_0)\delta\chi -\int_0^tdt'\dot\chi_0\frac{\csc(2\chi_0)}{\sqrt{\beta_0^2-\dot\chi_0^2}}\delta\dot\chi +\beta_0\delta\beta\int_0^tdt'\frac{\csc(2\chi_0)}{\sqrt{\beta_0^2-\dot\chi_0^2}}.\label{deltaxi}
\eeq
In the case of $\beta$-noise, we substitute the expressions for $\delta_\beta\chi(t)$ and $\delta_\beta\dot\chi(t)$ from Eq.~(\ref{deltahchifromPhi}) into Eq.~(\ref{deltaxi}) to obtain
\bea
\delta_\beta\xi(t)&=&-\delta\beta\frac{4}{\beta_0}\int_0^tdt'\sqrt{\beta_0^2-\dot\chi_0^2}\cot(2\chi_0)\csc(2\chi_0)\left\{\frac{1}{8}\sin(4\chi_0)+\hbox{Re}[e^{-2i\Phi(\chi_0)}G_\beta(\chi_0)]\right\}\nn\\
&&-\delta\beta\frac{1}{\beta_0}\int_0^tdt'\dot\chi_0^2\frac{\csc(2\chi_0)}{\sqrt{\beta_0^2-\dot\chi_0^2}}\left\{1+4\Phi'(\chi_0)\hbox{Im}[e^{-2i\Phi(\chi_0)}G_\beta(\chi_0)]\right\}+\beta_0\delta\beta\int_0^tdt'\frac{\csc(2\chi_0)}{\sqrt{\beta_0^2-\dot\chi_0^2}}.
\eea
Using the relation $\sqrt{\beta_0^2-\dot\chi_0^2}=\dot\chi_0\Phi'(\chi_0)\sin(2\chi_0)$, we can eliminate all of the square roots:
\bea
\delta_\beta\xi(t)&=&-\delta\beta\frac{4}{\beta_0}\int_0^tdt'\dot\chi_0\Phi'(\chi_0)\cot(2\chi_0)\left\{\frac{1}{8}\sin(4\chi_0)+\hbox{Re}[e^{-2i\Phi(\chi_0)}G_\beta(\chi_0)]\right\}\nn\\
&&-\delta\beta\frac{1}{\beta_0}\int_0^tdt'\dot\chi_0\frac{\csc^2(2\chi_0)}{\Phi'(\chi_0)}\left\{1+4\Phi'(\chi_0)\hbox{Im}[e^{-2i\Phi(\chi_0)}G_\beta(\chi_0)]\right\}+\beta_0\delta\beta\int_0^tdt'\frac{\csc^2(2\chi_0)}{\dot\chi_0\Phi'(\chi_0)}.
\eea
With the help of the relation
\beq
\dot\chi_0=\frac{\beta_0}{\sqrt{1+[\Phi'(\chi_0)]^2\sin^2(2\chi_0)}},
\eeq
we can write each term as an integral over $\chi$:
\bea
\delta_\beta\xi(t)&=&-\delta\beta\frac{4}{\beta_0}\int_0^{\chi_0(t)}d\chi\Phi'(\chi)\cot(2\chi)\left\{\frac{1}{8}\sin(4\chi)+\hbox{Re}[e^{-2i\Phi(\chi)}G_\beta(\chi)]\right\}\nn\\
&&-\delta\beta\frac{1}{\beta_0}\int_0^{\chi_0(t)}d\chi\frac{\csc^2(2\chi)}{\Phi'(\chi)}\left\{1+4\Phi'(\chi)\hbox{Im}[e^{-2i\Phi(\chi)}G_\beta(\chi)]\right\} +\delta\beta\frac{1}{\beta_0}\int_0^{\chi_0(t)}d\chi\frac{\csc^2(2\chi)}{\Phi'(\chi)}\left\{1+[\Phi'(\chi)]^2\sin^2(2\chi)\right\}\nn\\
&=&\delta\beta\frac{1}{\beta_0}\int_0^{\chi_0(t)}d\chi\Phi'(\chi)\left\{\sin^2(2\chi)-4\cot(2\chi)\hbox{Re}[e^{-2i\Phi(\chi)}G_\beta(\chi)]\right\} -\delta\beta\frac{4}{\beta_0}\int_0^{\chi_0(t)}d\chi\csc^2(2\chi)\hbox{Im}[e^{-2i\Phi(\chi)}G_\beta(\chi)]\nn\\&&\label{deltaxi2}
\eea
This result can be simplified dramatically if we perform an integration by parts on the last term:
\bea
\int_0^{\chi_0}d\chi\csc^2(2\chi)\hbox{Im}[e^{-2i\Phi(\chi)}G_\beta(\chi)]&=&-\frac{1}{2}\cot(2\chi)\hbox{Im}[e^{-2i\Phi(\chi)}G_\beta(\chi)]\bigg|_0^{\chi_0} -\int_0^{\chi_0}d\chi\cot(2\chi)\Phi'(\chi)\hbox{Re}[e^{-2i\Phi(\chi)}G_\beta(\chi)].\nn\\&&
\eea
The second term here precisely cancels a similar-looking term in Eq.~(\ref{deltaxi2}), leaving us with
\beq
\delta_\beta\xi(t)=\frac{\delta\beta}{\beta_0}\int_0^{\chi_0(t)}d\chi\Phi'(\chi)\sin^2(2\chi) +2\frac{\delta\beta}{\beta_0}\cot[2\chi_0(t)]\hbox{Im}\left\{e^{-2i\Phi[\chi_0(t)]}G_\beta[\chi_0(t)]\right\},
\eeq
where we have also used that
\beq
\lim_{\chi\to0}\cot(2\chi)\hbox{Im}[e^{-2i\Phi(\chi)}G_\beta(\chi)]=0.
\eeq
The requirement that $\delta_\beta\xi_f=0$ then yields the following condition:
\beq
\delta_\beta\xi_f=\frac{\delta\beta}{\beta_0}\int_0^{\chi_f}d\chi\Phi'(\chi)\sin^2(2\chi) +2\frac{\delta\beta}{\beta_0}\cot(2\chi_f)\hbox{Im}\left\{e^{-2i\Phi(\chi_f)}G_\beta(\chi_f)\right\}=0.
\eeq
However, the second term in this condition is already required to vanish according to the additional constraints we have derived earlier (see Eq.~(\ref{canceldherrori})), so the first term must also vanish independently:
\beq
\int_0^{\chi_f}d\chi\Phi'(\chi)\sin^2(2\chi)=0.\label{canceldherroriii}
\eeq
In summary, the following two constraints are necessary and sufficient for ensuring that the leading-order error in the evolution operator due to $\beta$-noise vanishes identically:
\bea
\sin(4\chi_f)+8e^{-2i\Phi(\chi_f)}\int_0^{\chi_f} d\chi \sin^2(2\chi)e^{2i\Phi(\chi)}&=&0,\nn\\
\int_0^{\chi_f}d\chi\Phi'(\chi)\sin^2(2\chi)&=&0.\label{betaconstraints}
\eea

\subsection{$\Omega(t)$ noise}

From the general form of the evolution operator, Eq.~(\ref{evolfromchi}), we see that in the case of a general single-axis driving field, the first-order error due to $\Omega$-noise will vanish only if the following three conditions on the first-order variations hold:
\beq
\delta_\epsilon\chi(t_f)=0,\qquad \delta_\epsilon\dot\chi(t_f)=0,\qquad \delta_\epsilon\xi(t_f)=0.\label{dJerror}
\eeq
In Sec.~\ref{sec:Jnoise}, we found the fluctations in $\chi(t)$ caused by fluctuations in $\Omega(t)$, Eq.~(\ref{deltachie}). Using the results of Sec.~\ref{sec:topology}, this can be rewritten as
\bea
\delta_\epsilon\chi(t)&=&\delta\epsilon\frac{1}{2\beta}\left\{\beta k\cos(2\Phi[\chi_0(t)])+\hbox{Im}\left[e^{-2i\Phi[\chi_0(t)]}\int_0^{\chi_0(t)}d\tilde\chi\sin(2\tilde\chi)\frac{\tilde g(\tilde\chi)}{F(\tilde\chi)}e^{2i\Phi(\tilde\chi)}\right]\right\},
\eea
where we have defined $\tilde g$ such that $g(t)=\tilde g(\chi_0(t))$. Defining the function
\beq
G_\epsilon(\chi)\equiv\int_0^{\chi}d\tilde\chi\sin(2\tilde\chi)\frac{\tilde g(\tilde\chi)}{F(\tilde\chi)}e^{2i\Phi(\tilde\chi)},
\eeq
and setting $k=0$ as is appropriate for fluctuations that originate from the control field $\Omega(t)$, this becomes
\bea
\delta_\epsilon\chi(t)&=&\delta\epsilon\frac{1}{2\beta}\hbox{Im}\left[e^{-2i\Phi[\chi_0(t)]}G_\epsilon[\chi_0(t)]\right],\nn\\
\delta_\epsilon\dot\chi(t)&=&-\delta\epsilon\frac{1}{\beta}\dot\chi_0(t)\Phi'[\chi_0(t)]\hbox{Re}\left[e^{-2i\Phi[\chi_0(t)]}G_\epsilon[\chi_0(t)]\right].\label{deltaechi}
\eea
As in the case of $\beta$-noise, we again want to avoid setting $\Phi'(\chi_f)=0$ since this will lead to an incomplete cancelation of the first-order error. We are then left with the following condition necessary for canceling the error:
\beq
G_\epsilon(\chi_f)=0.\label{cancelJnoise}
\eeq

We now turn to the third condition from Eq.~(\ref{dJerror}). For driving field noise, we may again use the first-order variation of $\xi(t)$, Eq.~(\ref{deltaxi}), but now drop the term involving $\delta\beta$:
\beq
\delta_\epsilon\xi(t)=-2\int_0^tdt'\sqrt{\beta_0^2-\dot\chi_0^2}\cot(2\chi_0)\csc(2\chi_0)\delta_\epsilon\chi -\int_0^tdt'\dot\chi_0\frac{\csc(2\chi_0)}{\sqrt{\beta_0^2-\dot\chi_0^2}}\delta_\epsilon\dot\chi.
\eeq
Using the expressions for $\delta_\epsilon\chi(t)$ and $\delta_\epsilon\dot\chi(t)$ given earlier in Eq.~(\ref{deltaechi}), this becomes
\beq
\delta_\epsilon\xi(t)=-\frac{\delta\epsilon}{\beta_0}\int_0^tdt'\sqrt{\beta_0^2-\dot\chi_0^2}\cot(2\chi_0)\csc(2\chi_0)\hbox{Im}[e^{-2i\Phi(\chi_0)}G_\epsilon(\chi_0)] +\frac{\delta\epsilon}{\beta_0}\int_0^tdt'\dot\chi_0^2\frac{\csc(2\chi_0)}{\sqrt{\beta_0^2-\dot\chi_0^2}}\Phi'(\chi_0)\hbox{Re}[e^{-2i\Phi(\chi_0)}G_\epsilon(\chi_0)].
\eeq
We next remove the square roots and turn the integrations over $t'$ into integrations over $\chi$ using $\sqrt{\beta_0^2-\dot\chi_0^2}=\dot\chi_0\Phi'(\chi_0)\sin(2\chi_0)$:
\beq
\delta_\epsilon\xi(t)=-\frac{\delta\epsilon}{\beta_0}\int_0^{\chi_0(t)}d\chi\Phi'(\chi)\cot(2\chi)\hbox{Im}[e^{-2i\Phi(\chi)}G_\epsilon(\chi)] +\frac{\delta\epsilon}{\beta_0}\int_0^{\chi_0(t)}d\chi\csc^2(2\chi)\hbox{Re}[e^{-2i\Phi(\chi)}G_\epsilon(\chi)].\label{deltaexi2}
\eeq
Integrating the second term by parts,
\bea
\int_0^{\chi_0}d\chi\csc^2(2\chi)\hbox{Re}[e^{-2i\Phi(\chi)}G_\epsilon(\chi)]&=&-\frac{1}{2}\cot(2\chi)\hbox{Re}[e^{-2i\Phi(\chi)}G_\epsilon(\chi)]\bigg|_0^{\chi_0} +\frac{1}{2}\int_0^{\chi_0}d\chi\cos(2\chi)\frac{{\tilde g}(\chi)}{F(\chi)}\nn\\ &&+\int_0^{\chi_0}d\chi\cot(2\chi)\Phi'(\chi)\hbox{Im}[e^{-2i\Phi(\chi)}G_\epsilon(\chi)],
\eea
we find that the third term here cancels the first term in Eq.~(\ref{deltaexi2}), leaving
\beq
\delta_\epsilon\xi(t)=\frac{\delta\epsilon}{2\beta_0}\int_0^{\chi_0(t)}d\chi\cos(2\chi)\frac{\tilde g(\chi)}{F(\chi)}-\frac{\delta\epsilon}{2\beta_0}\cot[2\chi_0(t)]\hbox{Re}\left\{e^{-2i\Phi[\chi_0(t)]}G_\epsilon[\chi_0(t)]\right\}.
\eeq
The end-point variation is then
\beq
\delta_\epsilon\xi_f=\frac{\delta\epsilon}{2\beta_0}\int_0^{\chi_f}d\chi\cos(2\chi)\frac{\tilde g(\chi)}{F(\chi)}-\frac{\delta\epsilon}{2\beta_0}\cot(2\chi_f)\hbox{Re}[e^{-2i\Phi(\chi_f)}G_\epsilon(\chi_f)].
\eeq
When we impose the earlier condition from Eq.~(\ref{cancelJnoise}), the second term vanishes, in turn requiring that the first term also vanishes independently. In summary, the following two constraints are necessary and sufficient for ensuring that the leading-order error in the evolution operator due to $\Omega$-noise vanishes identically:
\bea
\int_0^{\chi_f}d\chi\sin(2\chi)\tilde g(\chi)e^{2i\Phi(\chi)}\sqrt{1+[\Phi'(\chi)]^2\sin^2(2\chi)}&=&0,\nn\\
\int_0^{\chi_f}d\chi\cos(2\chi)\tilde g(\chi)\sqrt{1+[\Phi'(\chi)]^2\sin^2(2\chi)}&=&0.\label{Omegaconstraints}
\eea

\section{Constructing robust control fields}

Here, we collect all the preceding results and summarize the general approach to constructing robust control fields. These control fields are completely determined by a function $\Phi(\chi)$ which we are free to choose. We can make the initial evolution operator the identity matrix by imposing $\Phi(0)=0$ and $\Phi'(0)=0$. Choosing $\eta=-1$ in Eq.~(\ref{evolfromchi}) and focusing on the case of single-axis driving, we find that the target evolution operator is determined by the value of $\Phi$ and its first derivative at a chosen final value of $\chi$, which we denote by $\chi_f$:
\bea
U_{target}&=&\left(\begin{matrix}\cos\chi_fe^{i\xi_-(t_f)} & -i\sin\chi_fe^{-i\xi_+(t_f)} \cr -i\sin\chi_fe^{i\xi_+(t_f)} & \cos\chi_fe^{-i\xi_-(t_f)}\end{matrix}\right),\nn\\
\xi_\pm(t_f)&=&\Phi(\chi_f)\mp\frac{1}{2}\hbox{arcsec}\left(\sqrt{1+[\Phi'(\chi_f)]^2\sin^2(2\chi_f)}\right).
\eea
A driving field $\Omega(t)$ which generates this target evolution via the Hamiltonian $H=\Omega(t)\sigma_z+\beta\sigma_x$ can be found by performing the following integration
\beq
\beta t=\int_0^{\chi_0} d\chi\sqrt{1+[\Phi'(\chi)]^2\sin^2(2\chi)},
\eeq
inverting the result to obtain $\chi_0(t)$, and plugging this into
\beq
\Omega(\chi)=-\beta\frac{\Phi''(\chi)\sin(2\chi)+4\Phi'(\chi)\cos(2\chi)+2[\Phi'(\chi)]^3\sin^2(2\chi)\cos(2\chi)}{2\left\{1+[\Phi'(\chi)]^2\sin^2(2\chi)\right\}^{3/2}}.\label{bzfromPhi2}
\eeq
The most general way to ensure that the resulting pulse has finite duration is to require that the numerator of the expression for $\Omega(\chi)$, Eq.~(\ref{bzfromPhi2}), vanishes at $\chi=\chi_f$. We should think of this condition as a condition on the final value of the second derivative of $\Phi(\chi)$:
\beq
\Phi''(\chi_f)=-4\Phi'(\chi_f)\cot(2\chi_f)-2[\Phi'(\chi_f)]^3\sin(2\chi_f)\cos(2\chi_f).\label{Phippcondition}
\eeq
Viewing the condition this way is appropriate since we have already fixed $\chi_f$ and $\Phi'(\chi_f)$ in accordance with our desired $U_{target}$.

If we want the leading-order errors in the evolution to vanish, then we must constrain not only the boundary conditions of $\Phi(\chi)$ but its full behavior from $\chi=0$ to $\chi=\chi_f$ as well. In the case of $\beta$-noise, $\Phi(\chi)$ must satisfy
\bea
\sin(4\chi_f)+8e^{-2i\Phi(\chi_f)}\int_0^{\chi_f} d\chi \sin^2(2\chi)e^{2i\Phi(\chi)}&=&0,\nn\\
\int_0^{\chi_f}d\chi\Phi'(\chi)\sin^2(2\chi)&=&0.\label{betaconstraints2}
\eea
whereas in the case of $\Omega(t)$-noise, it must satisfy
\bea
\int_0^{\chi_f}d\chi\sin(2\chi)\tilde g(\chi)e^{2i\Phi(\chi)}\sqrt{1+[\Phi'(\chi)]^2\sin^2(2\chi)}&=&0,\nn\\
\int_0^{\chi_f}d\chi\cos(2\chi)\tilde g(\chi)\sqrt{1+[\Phi'(\chi)]^2\sin^2(2\chi)}&=&0,\label{Omegaconstraints2}
\eea
where the function $\tilde g(\chi)$ is determined by the precise nature of the noise one wishes to consider: $\delta\Omega=\tilde g(\chi_0(t))\delta\epsilon$. For example, pulse amplitude fluctuations are described by choosing $\tilde g(\chi)=\Omega(\chi)$. If one wishes to cancel errors due to both types of noise simultaneously, then $\Phi(\chi)$ must satisfy both sets of constraints.

The fact that the arbitrary time-dependence in the $\Omega$-noise fluctuation carries all the way through the calculation and leads to the appearance of the arbitrary function $\tilde g(\chi)$ in Eq.~(\ref{Omegaconstraints2}) suggests that the analysis could be extended to the case of non-static noise. In particular, one could consider performing a Gaussian path integral average of $\tilde g$, weighted by a nontrivial noise power spectrum. Since $\tilde g$ appears linearly inside a single integration in Eq.~(\ref{Omegaconstraints2}), it is likely that this average could be performed analytically using standard Gaussian integration techniques. This analysis will be carried out elsewhere.

A straightforward way to solve the above constraints on $\Phi(\chi)$, either Eqs.~(\ref{betaconstraints2}) or Eqs.~(\ref{Omegaconstraints2}), is to start with an ansatz for $\Phi(\chi)$ of the form
\beq
\Phi(\chi)=a_1\chi^2+a_2\chi^3+f(\chi),\label{Phiexample}
\eeq
where $a_1$ and $a_2$ are constants and $f$ is such that $f(0)=f'(0)=f'(\chi_f)=f''(\chi_f)=0$. We can set any desired $U_{target}$ by choosing $\chi_f$, $a_1$, and $a_2$ appropriately. We may then include additional parameters in $f(\chi)$ and tune these until the error-cancelation constraints are satisfied. For example, if we choose
\beq
f(\chi)=a_3\sin^3(\pi\chi/\chi_f)+a_4\sin^3(2\pi\chi/\chi_f),
\eeq
then we may freely tune $a_3$ and $a_4$ without changing $U_{target}$. This ansatz is used to construct the robust driving fields shown in Fig. 1 of the main text for the case of a driving field subject to noise in its amplitude.

Interestingly, it is also possible to solve the $\beta$-noise cancelation constraints analytically for a certain set of target rotations. In particular, we can solve the first constraint in Eq.~(\ref{betaconstraints2}) when $\chi_f=n\pi/4$ for some integer $n$ by choosing $\Phi(\chi)$ to have the general form
\beq
\Phi(\chi)=[4\chi-\sin(4\chi)+\lambda\zeta(4\chi-\sin(4\chi))]/n,
\eeq
where $\zeta(\theta+n\pi/2)=\zeta(\theta)$ is any periodic function with period $n\pi/2$, and $\lambda$ is an arbitrary real parameter. We can automatically solve the second constraint in Eq.~(\ref{betaconstraints2}) for any non-constant $\zeta(\theta)$ by setting $\lambda$ to
\beq
\lambda^{-1}=\frac{2}{3\pi n}\int_0^{n\pi}d\theta\sin\theta\zeta(\theta-\sin\theta).
\eeq
This solution gives rise to a class of target evolutions determined by $n$ and $\lambda\zeta'(n\pi)$; in particular, the phases in the evolution operator are
\beq
\xi_\pm(t_f)=n\pi\mp\frac{1}{2}\hbox{arcsec}\left(\sqrt{1+\frac{64}{n^2}\sin^6\left(\frac{n\pi}{2}\right)\left[1+\lambda\zeta'(n\pi)\right]^2}\right).
\eeq
If $n$ is an even integer but not a multiple of 4, then $U_{target}$ reduces to a $\pi$-rotation about the $x$ axis. If $n$ is a multiple of 4, then $U_{target}$ is proportional to the identity. In this case, the control field implements a smooth-pulse version of dynamical decoupling.

The fact that the $\beta$-noise constraints can be solved analytically greatly simplifies the problem in the case where one wishes to cancel both types of noise simultaneously. To do so, one could start with a more general ansatz for $\Phi$:
\beq
\Phi(\chi)=\frac{1}{n}\left[4\chi-\sin(4\chi)+\sum_k\lambda_k\zeta_k(4\chi-\sin(4\chi))\right],
\eeq
where all the $\zeta_k(\theta)$ are periodic functions of period $n\pi/2$. This ansatz automatically solves the $\beta$-noise constraints provided the $\lambda_k$ satisfy one linear constraint coming from the second condition in Eq.~(\ref{betaconstraints2}). The remaining $\lambda_k$ can then be tuned until the $\Omega$-noise constraints are also satisfied.

\section{Antisymmetric pulses}\label{sec:oddchi}
If $\chi$ is an odd function, $\chi(-t)=-\chi(t)$, which implies that $\Omega(-t)=-\Omega(t)$, then the total evolution operator which evolves the qubit from $t=-t_f$ to $t=t_f$ takes a particularly simple form which leads to fewer error-cancelation constraints that need to be satisfied. The antisymmetry of $\chi(t)$ immediately implies that the phases in the evolution operator enjoy the following property under time reversal:
\beq
\xi_\pm(-t_f)=\xi_\pm(t_f).
\eeq
Using this relation, it can be shown that the full evolution is described by the operator
\beq
U_{tot}=U(t_f)\sigma_zU(t_f)^\dag\sigma_z,\label{UtotfromUii}
\eeq
where $U(t_f)$ describes the evolution from $t=0$ to $t=t_f$. Eq.~(\ref{UtotfromUii}) leads to the expressions
\bea
U_{tot,11}&=&\cos(2\chi_f),\nn\\
U_{tot,12}&=&i\eta\sin(2\chi_f))e^{i\xi_-(t_f)-i\xi_+(t_f)},
\eea
from which we extract the following total evolution operator components for antisymmetric pulses:
\bea
\hbox{Re}[U_{tot,11}]&=&\cos(2\chi_f),\nn\\
\hbox{Im}[U_{tot,11}]&=&0,\nn\\
\hbox{Re}[U_{tot,12}]&=&\sqrt{1-\frac{\dot\chi_f^2}{\beta^2}}\sin(2\chi_f),\nn\\
\hbox{Im}[U_{tot,12}]&=&-\frac{\dot\chi_f}{\beta}\sin(2\chi_f).\label{oddchievolop}
\eea
This evolution operator corresponds to rotations about axes in the $xy$ plane. The axis and angle of rotation can be expressed in terms of $\chi_f$ and $\Phi'(\chi_f)$ according to
\beq
\phi=4\chi_f,\qquad \cos\theta=\frac{1}{\sqrt{1+[\Phi'(\chi_f)]^2\sin^2(2\chi_f)}},
\eeq
where $\theta$ is the angle between the axis of rotation (which lies in the $xy$ plane) and the $x$-axis, and $\phi$ is the angle of rotation. Thus finding a $\Phi(\chi)$ that implements a desired evolution amounts to imposing the following condition on the derivative of $\Phi(\chi)$ at $\chi_f=\phi/4$:
\beq
\Phi'(\phi/4)=\tan\theta\csc(\phi/2).
\eeq
Because $U_{tot}$ does not depend on $\xi(t_f)$ in the case of an antisymmetric pulse, we do not need to impose the error-cancelation constraint associated with fluctuations in $\xi(t_f)$. This means that we may ignore the second constraints in Eqs.~(\ref{betaconstraints2}) and (\ref{Omegaconstraints2}), and we need only solve the first constraints in each set to obtain robust antisymmetric control fields.

\section{Universal set of rotations}

To perform an arbitrary single qubit rotation, it suffices to be able to rotate by an arbitrary angle around one chosen axis, plus being able to do a $\pi$ rotation around an axis $\pi/4$ ($45$ degrees) apart from that chosen axis. The latter gate serves a similar role as the Hadamard gate which turns any rotation around one axis into a rotation with the same angle around an axis perpendicular to it.

In this section we present the parameters required to perform a set of rotations around an axis in the $x$-$y$ plane lying at an angle $\theta$ from the $x$-axis, namely
\begin{align}
R(\theta,\phi)=\exp\left[-i(\sin\theta\sigma_y+\cos\theta\sigma_x)\frac{\phi}{2}\right].
\end{align}

For each of the two cases discussed in the main text (pulses correcting driving field errors and $\delta\beta$ errors), we give pulses that implement rotations $R(5\pi/12,\pi)$ and $R(\pi/6,\phi)$ for a range of $\phi$. The two rotation axes are $\pi/4$ apart as required. Due to the strong nonlinearity of the problem, it is not practical to scan over the entire $[0,2\pi)$ range of $\phi$ continuously; we therefore demonstrate a set of rotations with angles between $\pi$ and $3\pi$ with step $0.2\pi$. Since our method is completely general and can take as many parameters as needed, it is straightforward to find the pulses for any single-qubit rotation.

\subsection{Pulses correcting driving field errors}

For pulses correcting driving field errors, we make use of the ansatz
\begin{align}
\Phi(\chi)=a_1\sin^2(a_2\chi)+a_3\sin^3\left(\frac{4\pi\chi}{\phi}\right)+a_4\sin^3\left(\frac{8\pi\chi}{\phi}\right).
\label{eq:djonly}
\end{align}
Parameters for the set of rotations discussed above are provided in Table~\ref{tab:para_djonly}.

\begin{table}[h]
 \centering
 \begin{tabular}{|c||c|c|c|c|c|c|}
 \hline
 $R(\theta,\phi)$ & $a_1$ & $a_2$ & $a_3$ & $a_4$ \\
 \hline
 \hline
 $R(5\pi/12,\pi)$ & 1.24402 & 3.00000 & 2.01146 & 1.26906\\
 \hline
 $R(\pi/6,\pi)$ & -0.577350 & 1.00000 & 2.29863 & 1.01756\\
 \hline
 $R(\pi/6,1.2\pi)$ & 0.273558 & 2.33333 & 2.37161 & 1.17764\\
 \hline
 $R(\pi/6,1.4\pi)$ & -0.237992 & 3.35888 & 2.30057 & 1.28314\\
 \hline
 $R(\pi/6,1.6\pi)$ & 1.72135 & 1.41514 & 2.70120 & 1.50434\\
 \hline
 $R(\pi/6,1.8\pi)$ & 9.96178 & 1.16814 & 0.829187 & 1.87861\\
 \hline
 $I$ & 1.00000 & 1.00000 & 0 & 1.99924\\
 \hline
 $R(\pi/6,2.2\pi)$ & 0.317272 & 7.94004 & 1.75940 & 3.25893\\
 \hline
 $R(\pi/6,2.4\pi)$ & -1.46213 & 1.54756 & -1.14086 & 1.33871\\
 \hline
 $R(\pi/6,2.6\pi)$ & 2.84789 & -0.676288 & 1.91457 & -0.791508\\
 \hline
 $R(\pi/6,2.8\pi)$ & 1.78210 & -0.568079 & -0.451979 & 0.271430\\
 \hline
 \end{tabular}
 \caption{Parameters of pulses correcting driving field errors for a set of rotations.}\label{tab:para_djonly}
\end{table}

\subsection{Pulses correcting $\delta\beta$-noise}

For pulses correcting $\delta\beta$-noise, we use the ansatz
\begin{align}
\Phi(\chi)=a_1\sin^2(a_2\chi)+a_3\sin^3\left(\frac{4N_3\pi\chi}{\phi}\right)+a_4\sin^3\left[\frac{\pi(\phi-4\chi)}{\phi}\right]+a_5\sin^3\left[\frac{\pi}{4}\chi(\phi-4\chi)\right],
\label{eq:djonly}
\end{align}
where $N_3$ must be an integer.
Parameters for the set of rotations discussed above are provided in Table~\ref{tab:para_dhonly}.

\begin{table}[h]
 \centering
 \begin{tabular}{|c||c|c|c|c|c|c|}
 \hline
 $R(\theta,\phi)$ & $a_1$ & $a_2$ & $a_3$ & $a_4$ & $a_5$ & $N_3$\\
 \hline
 \hline
 $R(5\pi/12,\pi)$ & 1.24402 & 3.00000 & -0.464647 & 1.42922 & 10.4704 & 1\\
 \hline
  $R(\pi/6,\pi)$ & -0.577350 & 1.00000 & -1.41354 & 0.480222 & 30.4015 & 1\\
 \hline
  $R(\pi/6,1.2\pi)$ & 0.273558 & 2.33333 & 0.727786 & 2.62177 & -3.27545 & 1\\
 \hline
   $R(\pi/6,1.4\pi)$ & -0.237992 & 3.35888 & 2.12682 & 4.02077 & -6.92089 & 1\\
 \hline
   $R(\pi/6,1.6\pi)$ & 1.72135 & 1.41514 & -0.725379 & 0.172736 & -0.885146 & 2\\
 \hline
   $R(\pi/6,1.8\pi)$ & 9.96178 & 1.16814 & -0.108777 & -1.01910 & -0.978449 & 1\\
 \hline
   $R(\pi/6,2.2\pi)$ & 0.317272 & 7.94004 & -3.12414 & 0.803819 & 2.94582 & 2\\
 \hline
   $R(\pi/6,2.4\pi)$ & 5.45961 & 0.770847 & 2.74485 & 8.04491 & 4.95240 & 1\\
 \hline
   $R(\pi/6,2.6\pi)$ & 2.84789 & -0.676289 & 4.24974 & 5.15883 & 4.27094 & 1\\
 \hline
   $R(\pi/6,2.8\pi)$ & -0.598662 & 1.19958 & -1.14903 & -0.24894 & 0.665241 & 1\\
 \hline
  \end{tabular}
 \caption{Parameters of pulses correcting $\delta\beta$-noise for a set of rotations.}\label{tab:para_dhonly}
\end{table}

\section{Construction of ordinary pulses and definition of fidelity}

In Figs. 2 and 3 of the main text, we verify that the first-order error in the evolution operator has been canceled by comparing the infidelity of the designed pulse with that of a square pulse sequence that implements the same target evolution but which has not been designed to combat errors. In the case of single-axis driving, we can systematically construct a square pulse sequence that generates any target evolution operator using the following general form involving two identical square pulses:
\beq
U_{target}=R_{\beta_0}(0;t_c)R_{\beta_0}(\beta_0;\tau)R_{\beta_0}(0;t_b)R_{\beta_0}(\beta;\tau)R_{\beta_0}(0;t_a),
\eeq
where $R_\beta(\Omega;t)\equiv e^{-it(\Omega\sigma_z+\beta\sigma_x)}$ and $\tau=\pi/(2\sqrt{2}\beta_0)$. Given $U_{target}$, we can solve this equation numerically to obtain the parameters $t_a$, $t_b$, $t_c$. In Figs. 2c and 3c of the main text, we use the definition of infidelity as in Ref.~\onlinecite{Bowdrey_PLA02}:
\beq
\hbox{infidelity}=\frac{1}{2}-\frac{1}{12}\sum_{x,y,z}\hbox{Tr}\left\{U_{target}\sigma_jU_{target}^\dagger U(t_f)\sigma_j U^\dagger(t_f)\right\}.
\eeq
Here, $U(t_f)$ is the actual evolution operator including the errors to all orders. In the case of $\beta$-noise for example, the evolution corresponding to the uncorrected pulses is given simply by
\beq
U(t_f)=R_{\beta_0+\delta\beta}(0;t_c)R_{\beta_0+\delta\beta}(\beta_0;\tau)R_{\beta_0+\delta\beta}(0;t_b)R_{\beta_0+\delta\beta}(\beta;\tau)R_{\beta_0+\delta\beta}(0;t_a),
\eeq
while $U(t_f)$ for the designed pulses is computed by solving the Schr\"odinger equation numerically with Hamiltonian $H=\Omega(t)\sigma_z+(\beta_0+\delta\beta)\sigma_x$ for each value of the error strength, $\delta\beta$.


\begin{thebibliography}{10}
\expandafter\ifx\csname url\endcsname\relax
  \def\url#1{\texttt{#1}}\fi
\expandafter\ifx\csname urlprefix\endcsname\relax\def\urlprefix{URL }\fi
\providecommand{\bibinfo}[2]{#2}
\providecommand{\eprint}[2][]{\url{#2}}

\bibitem{Maune_Nature12}
\bibinfo{author}{Maune, B.~M.} \emph{et~al.}
\newblock \bibinfo{title}{Coherent singlet-triplet oscillations in a
  silicon-based double quantum dot}.
\newblock \emph{\bibinfo{journal}{Nature}} \textbf{\bibinfo{volume}{481}},
  \bibinfo{pages}{7381} (\bibinfo{year}{2012}).

\bibitem{Shulman_Science12}
\bibinfo{author}{Shulman, M.~D.} \emph{et~al.}
\newblock \bibinfo{title}{Demonstration of entanglement of electrostatically
  coupled singlet-triplet qubits}.
\newblock \emph{\bibinfo{journal}{Science}} \textbf{\bibinfo{volume}{336}},
  \bibinfo{pages}{202} (\bibinfo{year}{2012}).

\bibitem{Petta_Science05}
\bibinfo{author}{Petta, J.~R.} \emph{et~al.}
\newblock \bibinfo{title}{Coherent manipulation of coupled electron spins in
  semiconductor quantum dots}.
\newblock \emph{\bibinfo{journal}{Science}} \textbf{\bibinfo{volume}{309}},
  \bibinfo{pages}{2180--2184} (\bibinfo{year}{2005}).

\bibitem{vanderSar_Nature12}
\bibinfo{author}{van~der Sar, T.} \emph{et~al.}
\newblock \bibinfo{title}{Decoherence-protected quantum gates for a hybrid
  solid-state spin register}.
\newblock \emph{\bibinfo{journal}{Nature}} \textbf{\bibinfo{volume}{484}},
  \bibinfo{pages}{82--86} (\bibinfo{year}{2012}).

\bibitem{Pop_Nature14}
\bibinfo{author}{Pop, I.~M.} \emph{et~al.}
\newblock \bibinfo{title}{Coherent suppression of electromagnetic dissipation
  due to superconducting quasiparticles}.
\newblock \emph{\bibinfo{journal}{Nature}} \textbf{\bibinfo{volume}{508}},
  \bibinfo{pages}{369--372} (\bibinfo{year}{2014}).

\bibitem{Shor_PRA95}
\bibinfo{author}{Shor, P.~W.}
\newblock \bibinfo{title}{Scheme for reducing decoherence in quantum computer
  memory}.
\newblock \emph{\bibinfo{journal}{Phys.\ Rev.\ A}}
  \textbf{\bibinfo{volume}{52}}, \bibinfo{pages}{R2493(R)}
  (\bibinfo{year}{1995}).

\bibitem{Hahn_PR50}
\bibinfo{author}{Hahn, E.~L.}
\newblock \bibinfo{title}{Spin echoes}.
\newblock \emph{\bibinfo{journal}{Phys. Rev.}} \textbf{\bibinfo{volume}{80}},
  \bibinfo{pages}{580} (\bibinfo{year}{1950}).

\bibitem{Carr_Purcell}
\bibinfo{author}{Carr, H.~Y.} \& \bibinfo{author}{Purcell, E.~M.}
\newblock \bibinfo{title}{Effects of diffusion on free precession in nuclear
  magnetic resonance experiments}.
\newblock \emph{\bibinfo{journal}{Phys. Rev.}} \textbf{\bibinfo{volume}{94}},
  \bibinfo{pages}{640} (\bibinfo{year}{1954}).

\bibitem{Wimperis_JMR94}
\bibinfo{author}{Wimperis, S.}
\newblock \bibinfo{title}{Broadband, narrowband and passband composite pulses
  for use in advanced {NMR} experiments}.
\newblock \emph{\bibinfo{journal}{J.\ Magn.\ Reson.\ B}}
  \textbf{\bibinfo{volume}{109}}, \bibinfo{pages}{221} (\bibinfo{year}{1994}).

\bibitem{Merrill_Wiley14}
\bibinfo{author}{Merrill, J.~T.} \& \bibinfo{author}{Brown, K.~R.}
\newblock \bibinfo{title}{in Progress in compensating pulse sequences for quantum
  computation, in quantum information and computation for chemistry: Advances
  in chemical physics,} \bibinfo{pages}{Vol. 154 (ed. Kais, S.) (John Wiley \&
  Sons Inc.,} \bibinfo{year}{2014}).

\bibitem{Uhrig_PRL07}
\bibinfo{author}{Uhrig, G.~S.}
\newblock \bibinfo{title}{Keeping a quantum bit alive by optimized $\pi$-pulse
  sequences}.
\newblock \emph{\bibinfo{journal}{Phys.\ Rev.\ Lett.}}
  \textbf{\bibinfo{volume}{98}}, \bibinfo{pages}{100504}
  (\bibinfo{year}{2007}).

\bibitem{Witzel_PRL07}
\bibinfo{author}{Witzel, W.~M.} \& \bibinfo{author}{{Das Sarma}, S.}
\newblock \bibinfo{title}{Multiple-pulse coherence enhancement of solid state
  spin qubits}.
\newblock \emph{\bibinfo{journal}{Phys.\ Rev.\ Lett.}}
  \textbf{\bibinfo{volume}{98}}, \bibinfo{pages}{077601}
  (\bibinfo{year}{2007}).

\bibitem{Du_Nature09}
\bibinfo{author}{Du, J.} \emph{et~al.}
\newblock \bibinfo{title}{{Preserving electron spin coherence in solids by
  optimal dynamical decoupling}}.
\newblock \emph{\bibinfo{journal}{{Nature}}} \textbf{\bibinfo{volume}{{461}}},
  \bibinfo{pages}{{1265--1268}} (\bibinfo{year}{{2009}}).

\bibitem{Goelman_JMR89}
\bibinfo{author}{Goelman, G.}, \bibinfo{author}{Vega, S.} \&
  \bibinfo{author}{Zax, D.~B.}
\newblock \bibinfo{title}{Squared amplitude-modulated composite pulses}.
\newblock \emph{\bibinfo{journal}{J.\ Magn.\ Reson.}}
  \textbf{\bibinfo{volume}{81}}, \bibinfo{pages}{423} (\bibinfo{year}{1989}).

\bibitem{Khodjasteh_PRL10}
\bibinfo{author}{Khodjasteh, K.}, \bibinfo{author}{Lidar, D.~A.} \&
  \bibinfo{author}{Viola, L.}
\newblock \bibinfo{title}{Arbitrarily accurate dynamical control in open
  quantum systems}.
\newblock \emph{\bibinfo{journal}{Phys.\ Rev.\ Lett.}}
  \textbf{\bibinfo{volume}{104}}, \bibinfo{pages}{090501}
  (\bibinfo{year}{2010}).

\bibitem{Wang_NatComm12}
\bibinfo{author}{Wang, X.} \emph{et~al.}
\newblock \bibinfo{title}{Composite pulses for robust universal control of
  singlet-triplet qubits}.
\newblock \emph{\bibinfo{journal}{Nature Communications}}
  \textbf{\bibinfo{volume}{3}}, \bibinfo{pages}{997} (\bibinfo{year}{2012}).

\bibitem{Kestner_PRL13}
\bibinfo{author}{Kestner, J.~P.}, \bibinfo{author}{Wang, X.},
  \bibinfo{author}{Bishop, L.~S.}, \bibinfo{author}{Barnes, E.} \&
  \bibinfo{author}{{Das Sarma}, S.}
\newblock \bibinfo{title}{Noise-resistant control for a spin qubit array}.
\newblock \emph{\bibinfo{journal}{Phys.\ Rev.\ Lett.}}
  \textbf{\bibinfo{volume}{110}}, \bibinfo{pages}{140502}
  (\bibinfo{year}{2013}).

\bibitem{Wang_PRA14}
\bibinfo{author}{Wang, X.}, \bibinfo{author}{Bishop, L.~S.},
  \bibinfo{author}{Barnes, E.}, \bibinfo{author}{Kestner, J.~P.} \&
  \bibinfo{author}{{Das Sarma}, S.}
\newblock \bibinfo{title}{Robust quantum gates for singlet-triplet spin qubits
  using composite pulses}.
\newblock \emph{\bibinfo{journal}{Phys.\ Rev.\ A}}
  \textbf{\bibinfo{volume}{89}}, \bibinfo{pages}{022310}
  (\bibinfo{year}{2014}).

\bibitem{Kabytayev_PRA14}
\bibinfo{author}{Kabytayev, C.} \emph{et~al.}
\newblock \bibinfo{title}{Robustness of composite pulses to time-dependent
  control noise}.
\newblock \emph{\bibinfo{journal}{Phys.\ Rev.\ A}}
  \textbf{\bibinfo{volume}{90}}, \bibinfo{pages}{012316}
  (\bibinfo{year}{2014}).

\bibitem{Khodjasteh_PRA12}
\bibinfo{author}{Khodjasteh, K.}, \bibinfo{author}{Bluhm, H.} \&
  \bibinfo{author}{Viola, L.}
\newblock \bibinfo{title}{Automated synthesis of dynamically corrected quantum
  gates}.
\newblock \emph{\bibinfo{journal}{Phys.\ Rev.\ A}}
  \textbf{\bibinfo{volume}{86}}, \bibinfo{pages}{042329}
  (\bibinfo{year}{2012}).

\bibitem{Soare_NP14}
\bibinfo{author}{Soare, A.} \emph{et~al.}
\newblock \bibinfo{title}{Experimental noise filtering by quantum control}.
\newblock \emph{\bibinfo{journal}{Nature\ Phys.}}
  \textbf{\bibinfo{volume}{10}}, \bibinfo{pages}{825} (\bibinfo{year}{2014}).

\bibitem{Fauseweh_PRA12}
\bibinfo{author}{Fauseweh, B.}, \bibinfo{author}{Pasini, S.} \&
  \bibinfo{author}{Uhrig, G.~S.}
\newblock \bibinfo{title}{Frequency modulated pulses for quantum bits coupled
  to time-dependent baths}.
\newblock \emph{\bibinfo{journal}{Phys.\ Rev.\ A}}
  \textbf{\bibinfo{volume}{85}}, \bibinfo{pages}{022310}
  (\bibinfo{year}{2012}).

\bibitem{Noebauer_arXiv14}
\bibinfo{author}{Noebauer, T.} \emph{et~al.}
\newblock \bibinfo{title}{Smooth optimal quantum control for robust solid state spin magnetometry}.
\newblock \bibinfo{pages}{arXiv:1412.5051} (\bibinfo{year}{2014}).

\bibitem{Barnes_PRL12}
\bibinfo{author}{Barnes, E.} \& \bibinfo{author}{{Das Sarma}, S.}
\newblock \bibinfo{title}{Analytically solvable driven time-dependent two-level
  quantum systems}.
\newblock \emph{\bibinfo{journal}{Phys.\ Rev.\ Lett.}}
  \textbf{\bibinfo{volume}{109}}, \bibinfo{pages}{060401}
  (\bibinfo{year}{2012}).

\bibitem{Levy.02}
\bibinfo{author}{Levy, J.}
\newblock \bibinfo{title}{Universal quantum computation with spin-$1/2$ pairs
  and heisenberg exchange}.
\newblock \emph{\bibinfo{journal}{Phys.\ Rev.\ Lett.}}
  \textbf{\bibinfo{volume}{89}}, \bibinfo{pages}{147902}
  (\bibinfo{year}{2002}).

\bibitem{Wu_PNAS14}
\bibinfo{author}{Wu, X.} \emph{et~al.}
\newblock \bibinfo{title}{Two-axis control of a singlet-triplet qubit with an
  integrated micromagnet}.
\newblock \emph{\bibinfo{journal}{Proc.\ Natl.\ Acad.\ Sci.}}
  \textbf{\bibinfo{volume}{111}}, \bibinfo{pages}{11938}
  (\bibinfo{year}{2014}).

\bibitem{Economou_PRL07}
\bibinfo{author}{Economou, S.~E.} \& \bibinfo{author}{Reinecke, T.~L.}
\newblock \bibinfo{title}{Theory of fast optical spin rotation in a quantum dot
  based on geometric phases and trapped states}.
\newblock \emph{\bibinfo{journal}{Phys.\ Rev.\ Lett.}}
  \textbf{\bibinfo{volume}{99}}, \bibinfo{pages}{217401}
  (\bibinfo{year}{2007}).

\bibitem{Greilich_NP09}
\bibinfo{author}{Greilich, A.} \emph{et~al.}
\newblock \bibinfo{title}{Ultrafast optical rotations of electron spins in
  quantum dots}.
\newblock \emph{\bibinfo{journal}{Nat. Phys.}} \textbf{\bibinfo{volume}{5}},
  \bibinfo{pages}{262} (\bibinfo{year}{2009}).

\bibitem{Pla_Nature12}
\bibinfo{author}{Pla, J.~J.} \emph{et~al.}
\newblock \bibinfo{title}{A single-atom electron spin qubit in silicon}.
\newblock \emph{\bibinfo{journal}{Nature}} \textbf{\bibinfo{volume}{489}},
  \bibinfo{pages}{541} (\bibinfo{year}{2012}).

\bibitem{Rong.14}
\bibinfo{author}{Rong, X.} \emph{et~al.}
\newblock \bibinfo{title}{Implementation of dynamically corrected gates on a
  single electron spin in diamond}.
\newblock \emph{\bibinfo{journal}{Phys. Rev. Lett.}}
  \textbf{\bibinfo{volume}{112}}, \bibinfo{pages}{050503}
  (\bibinfo{year}{2014}).

\bibitem{Barnes_PRA13}
\bibinfo{author}{Barnes, E.}
\newblock \bibinfo{title}{Analytically solvable two-level quantum systems and
  landau-zener interferometry}.
\newblock \emph{\bibinfo{journal}{Phys.\ Rev.\ A}}
  \textbf{\bibinfo{volume}{88}}, \bibinfo{pages}{013818}
  (\bibinfo{year}{2013}).

\bibitem{Garanin_EPL02}
\bibinfo{author}{Garanin, D.~A.} \& \bibinfo{author}{Schilling, R.}
\newblock \bibinfo{title}{Inverse problem for the landau-zener effect}.
\newblock \emph{\bibinfo{journal}{Europhys.\ Lett.}}
  \textbf{\bibinfo{volume}{59}}, \bibinfo{pages}{7} (\bibinfo{year}{2002}).

\bibitem{Kane_Nature98}
\bibinfo{author}{Kane, B.~E.}
\newblock \bibinfo{title}{A silicon-based nuclear spin quantum computer}.
\newblock \emph{\bibinfo{journal}{Nature}} \textbf{\bibinfo{volume}{393}},
  \bibinfo{pages}{133--137} (\bibinfo{year}{1998}).

\bibitem{Morello_Nature10}
\bibinfo{author}{Morello, A.} \emph{et~al.}
\newblock \bibinfo{title}{Single-shot readout of an electron spin in silicon}.
\newblock \emph{\bibinfo{journal}{Nature}} \textbf{\bibinfo{volume}{467}},
  \bibinfo{pages}{687--691} (\bibinfo{year}{2010}).

\bibitem{Tyryshkin_NatMat11}
\bibinfo{author}{Tyryshkin, A.~M.} \emph{et~al.}
\newblock \bibinfo{title}{Electron spin coherence exceeding seconds in
  high-purity silicon}.
\newblock \emph{\bibinfo{journal}{Nat. Mater.}} \textbf{\bibinfo{volume}{11}},
  \bibinfo{pages}{143} (\bibinfo{year}{2011}).

\bibitem{Wolfowicz_NatNano13}
\bibinfo{author}{Wolfowicz, G.} \emph{et~al.}
\newblock \bibinfo{title}{Atomic clock transitions in silicon-based spin
  qubits}.
\newblock \emph{\bibinfo{journal}{Nat. Nanotech.}}
  \textbf{\bibinfo{volume}{8}}, \bibinfo{pages}{561} (\bibinfo{year}{2013}).

\bibitem{Pla.13}
\bibinfo{author}{Pla, J.~J.} \emph{et~al.}
\newblock \bibinfo{title}{High-fidelity readout and control of a nuclear spin
  qubit in silicon}.
\newblock \emph{\bibinfo{journal}{Nature}} \textbf{\bibinfo{volume}{496}},
  \bibinfo{pages}{334--338} (\bibinfo{year}{2013}).

\bibitem{Grinolds_NatPhys11}
\bibinfo{author}{Grinolds, M.~S.} \emph{et~al.}
\newblock \bibinfo{title}{Quantum control of proximal spins using nanoscale
  magnetic resonance imaging}.
\newblock \emph{\bibinfo{journal}{Nat.\ Phys.}} \textbf{\bibinfo{volume}{7}},
  \bibinfo{pages}{687--692} (\bibinfo{year}{2011}).

\end{thebibliography}
\end{document}